\documentclass[aps,prb,twocolumn,amsmath,amssymb,revsymb]{revtex4-1}%{preprintnumbers,showpacs,revtex4-1}
\usepackage{graphicx}% Include figure files
\usepackage{color} % color online
\usepackage{dcolumn}% Align table columns on decimal point
\usepackage{bm}% bold math
\usepackage{braket}% nice bras and kets
\usepackage{txfonts}

\definecolor{gray}{rgb}{0.5,0.5,0.5}

\newcommand{\fo}{\mathfrak{f}_{o}}

\providecommand{\h}{\hbar}
\providecommand{\p}{\partial}

\newcommand{\kb}{k_{\textsc{b}}^{}}
\newcommand{\om}{\omega}

\begin{document}

\title{Statistical theory of relaxation of high energy electrons in quantum Hall edge states}

\preprint{}

\author{Anders~Mathias~Lunde$^{1}$}\email[Corresponding author: ]{lunan@nbi.ku.dk}
\author{Simon~E.~Nigg$^{2}$}
\affiliation{$^{1}$Center for Quantum Devices, Niels Bohr Institute, University of Copenhagen, DK-2100 Copenhagen,  Denmark}
\affiliation{$^{2}$Department of Physics, University of Basel, Klingelbergstrasse 82, 4056 Basel, Switzerland}
\date{\today}

\begin{abstract}
We investigate theoretically the energy exchange between electrons of two co-propagating, out-of-equilibrium edge states with opposite spin polarization in the integer quantum Hall regime. A quantum dot tunnel-coupled to one of the edge states locally injects electrons at high energy. Thereby a narrow peak in the energy distribution is created at high energy above the Fermi level. A second downstream quantum dot performs an \emph{energy resolved} measurement of the electronic distribution function.  By varying the distance between the two dots, we are able to follow \emph{every step} of the energy exchange and relaxation between the edge states --- even analytically under certain conditions. In the absence of translational invariance along the edge, e.g.~due to the presence of disorder, energy can be exchanged by non-momentum conserving two-particle collisions. For weakly broken translational invariance, we show that the relaxation is described by coupled Fokker-Planck equations.  From these we find that relaxation of the injected electrons can be understood statistically as a generalized drift-diffusion process in energy space for which we determine the drift-velocity and the \emph{dynamical} diffusion parameter. Finally, we provide a physically appealing picture in terms of individual edge state heating as a result of the relaxation of the
injected electrons. 
\end{abstract}

\maketitle

\section{Introduction\label{sec:introduction}}

\begin{figure}[h!]
 \includegraphics[width=0.98\columnwidth]{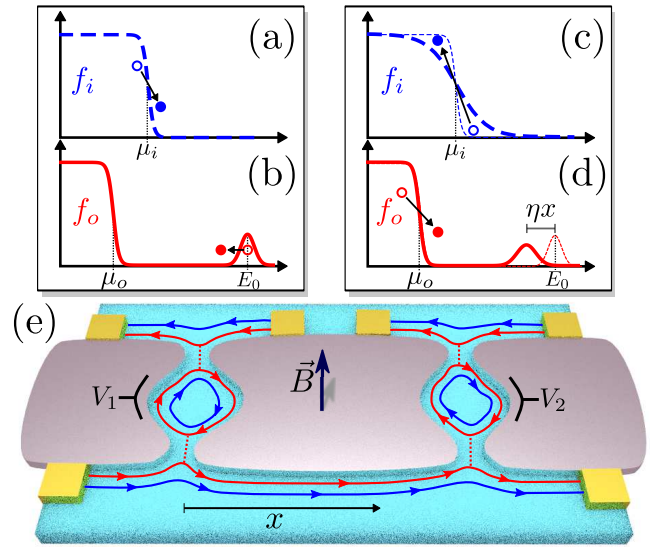}
\caption{(Color online) 
(e): Setup for the energy resolved injection and detection of electrons along the outer edge state (ES) (red lines). The QDs act as energy filters. The energy level of the left QD determines the (average) injection energy $E_0$ and can be tuned by the gate voltage $V_1$. By varying the gate voltage $V_2$, the energy dependence of the outer ES's electronic distribution $f_o$ can be measured. The inner ES's distribution $f_i$ can also be measured by opening the right QD completely for the outer ES. Thereby, the changes of both electronic distributions can be measured at a distance $x$ from the creation point. (a)-(d): Cartoon of the elementary energy exchange processes due to the inter-ES interaction: An electron injected into the outer ES looses energy (b), while an electron in the inner (blue) ES gains energy (a). This process smears out the Fermi sea of the inner ES, or intuitively, heats the inner ES. Subsequently, an electron in the inner (smeared) ES looses energy (c), while an electron in the Fermi sea of the outer ES gains energy (d). This sequence of energy exchange processes causes the energy distribution of the injected electrons (EDIE) to move downwards in energy at constant speed $\eta$ and broaden, while the Fermi seas of the inner and outer ESs smear (i.e. heat up).}
\label{fig:1}
\end{figure}

The integer quantum Hall effect is beautifully explained by a non-interacting theory of electrons propagating chirally along the edges of the two-dimensional electron gas in a perpendicular magnetic field.\cite{Halperin-PRB-1982,Buttiker-PRB-1988}   Even though multiple edge states propagate side-by-side along a single edge, their chirality prevents electronic backscattering, which is at the root of the extraordinary experimental precision of the quantized Hall conductivity.\cite{Buttiker-PRB-1988} Moreover, already early experiments demonstrated selective population and carrier detection in \emph{different} edge states on the same edge and thereby the first step was taken towards studying the dynamics between the edge states.\cite{Alphenaar-et-al-PRL1990,van-Wees-et-al-PRL-1989,Komiyama-PRB-1989,Muller-Khaetskii-et-al-PRB-1992,Khaetskii-PRB-1992} Interestingly, these studies showed that  forward scattering between individual edge states is strongly suppressed on very long distances -- but, unlike backscattering, not entirely absent -- and, moreover, depends on the spin-polarization of the involved edge states. If several edge states have the same spin-polarization, then inter edge state (ES) electron transfer was originally measured to take place on a scale of around 50 to 100 $\mu$m.\cite{Alphenaar-et-al-PRL1990,van-Wees-et-al-PRL-1989,Komiyama-PRB-1989} Newer experiments show a strong dependence of scattering between spin-degenerate ESs on disorder along the edge.\cite{Paradiso-1-PRB-2011,Paradiso-2-PRB-2011} However, if only two ESs with \emph{opposite} spin-polarization co-propagate (i.e. at filling factor $\nu=2$), then inter-ES electron scattering happens on even longer scales of up to a mm, \cite{Muller-Khaetskii-et-al-PRB-1992} since it is limited by the necessity of an electronic spin-flip.\cite{Khaetskii-PRB-1992}

Recently though, experimental evidence of energy exchange, {\em without electron exchange}, between co-propagating ESs was found in a quantum Hall system of filling factor $\nu=2$. Here one of the two ESs was intentionally brought out of equilibrium via a quantum point contact (QPC),\cite{Altimiras-Nat-phys-2010,le-Sueur-Altimiras-PRL-2010,Altimiras-2010a,Otsuka-et-al-2014} and subsequently an energy resolved measurement of the entire distribution function was performed by a single level quantum dot (QD) downstream from the QPC.\cite{Altimiras-Nat-phys-2010,le-Sueur-Altimiras-PRL-2010,Altimiras-2010a,Otsuka-PRB-2010}  This showed that the QPC created a step in the energy distribution function, in accordance with the non-interacting B\"uttiker scattering theory,\cite{Buttiker-PRB-1988} and the step did not change measurably on a distance of $\sim1 \mu$m.\cite{Altimiras-Nat-phys-2010} However, on longer distances of $\sim10 -  30 \mu$m,  a Fermi function with elevated temperature (compared to the fridge temperature) and shifted chemical potential was measured indicating energy exchange \emph{without} electron exchange between the ESs.\cite{le-Sueur-Altimiras-PRL-2010}  The energy exchange is most likely due to inter-ES electron-electron interactions.\cite{le-Sueur-Altimiras-PRL-2010,Lunde-2010,Degiovanni-PRB-2010}  Interestingly, it was also found that when one ES was forced to form a short closed loop, then energy exchange between the ESs was strongly suppressed.\cite{Altimiras-2010a}  

Another recent experiment\cite{Prokudina-et-al-PRL-2014} studies relaxation of two \emph{counter}-propagating non-equilibrium ESs by having an ES on either side of a thin gate in the quantum Hall regime of filling factor $\nu=1$ as previously discussed theoretically.\cite{Prokudina-et-al-JETP-2012} A different experiment also at $\nu=1$ found that heat only propagates along the electronic transport direction of the ES.\cite{Granger-PRL-2009} This experiment also revealed that hot electrons injected locally cooled down, while traveling along a single ES on a distance of roughly  60 $\mu$m.    

All these thorough experimental studies prompted various rather different theoretical works aimed at understanding the mechanism of energy relaxation in the integer quantum Hall regime.\cite{Lunde-2010,Degiovanni-PRB-2010,Kovrizhin-Chalker-PRB-2011,Prokudina-et-al-JETP-2012,Karzig-Levchenko-Glazman-Oppen-NJP-2012,Kovrizhin-Chalker-PRL-2012,Levkivskyi-Sukhorukov-PRB-2012,Levkivskyi-Sukhorukov-PRL-2012,Ngo-Dinh-et-al-PRB-2013,Nigg-PhysicaE-2016} Some of these works assume the co-propagating ESs to be translation invariant and their low-energy physics to be described by chiral Luttinger liquid (LL) theory.\cite{Degiovanni-PRB-2010,Kovrizhin-Chalker-PRB-2011,Kovrizhin-Chalker-PRL-2012,Levkivskyi-Sukhorukov-PRB-2012,Levkivskyi-Sukhorukov-PRL-2012,Ngo-Dinh-et-al-PRB-2013} Interestingly, the standard (chiral) LL model cannot strictly speaking account for energy relaxation due to its integrability.\cite{Takei-Milletari-Rosenow-PRB-2010,Micklitz-Jerome-Matveev-PRB-2010,Karzig-Levchenko-Glazman-Oppen-NJP-2012,Ngo-Dinh-et-al-PRB-2013,Slobodeniuk-PRB-2016} In order to study relaxation within a LL model one needs to circumvent this somehow.  The LL model first put forward by Degiovanni \emph{et al.} \cite{Degiovanni-PRB-2010} focused on energy exchange as a result of scattering of collective plasmon excitations in a region of finite size combined with a phenomenological model of the plasmon distribution due to the QPC. In Refs.~\onlinecite{Kovrizhin-Chalker-PRB-2011,Ngo-Dinh-et-al-PRB-2013}, the QPC is emulated by assuming an initial out-of-equilibrium step momentum distribution function and then its time-evolution is studied. While the effect of the QPC is easy to describe in a fermionic language, it is in general harder to do so within a bosonisation approach. Despite tremendous progress, a new experiment\cite{Tewari-PRB-2016} suggests that we still lack a complete theoretical understanding of interaction induced relaxation and decoherence in the integer quantum Hall regime.

In this and a previous work by the authors\cite{Lunde-2010} as well as in Ref.~\onlinecite{Prokudina-et-al-JETP-2012}, a different approach is used by explicitly taking into account the non-translation invariant nature of ESs.  This is due to the fact that ESs follow the equipotential lines, which are deformed by the presence of impurities and variations of the edge confinement potential.\cite{Buttiker-PRB-1988} Consequently, {\em non-momentum conserving} two-particle electron-electron interactions are present and we describe the relaxation in terms of a kinetic Boltzmann equation. In this model, non-equilibrium ES distributions are naturally predicted to relax towards Fermi distributions. In passing, we note that three-particle interactions,\cite{Lunde-PRB-2007} relevant for a translation invariant edge, have also been considered recently within a Fermi-liquid picture.\cite{Karzig-Levchenko-Glazman-Oppen-NJP-2012} Similar models for other quasi-one-dimensional systems have also recently been studied.\cite{Lunde-PRL-2006,Lunde-PRB-2007,Jerome-Micklitz-Matveev-PRL-2009,Lunde-NJP-09,Levchenko-Micklitz-Rech-Matveev-PRB-2010,Karzig-PRL-2010,Micklitz-Jerome-Matveev-PRB-2010,Huang-Gumbs-2010,Micklitz-Levchenko-PRL-2011,Dmitriev-Gornyi-Polyakov-PRB-2012,Nagaev-Sergeeva-PRB-2012,Micklitz-Levchenko-Rosch-PRL-2012,Imambekov-Schmidt-Glazman-RMP-2012,Rieder-et-al-PRB-2014}

Here we discuss the relaxation of high energy electrons injected locally by a side-coupled QD into one of   two ESs ($\nu=2$). This gives a narrow peak in the ES's energy distribution high above the Fermi level [Fig.~\ref{fig:1}(b)]. A second downstream QD measures the distribution function versus energy, see Fig.~\ref{fig:1}(e).~We analyze the manner in which the energy distribution of the injected electrons (EDIE) changes as the electrons relax towards the Fermi level. As a consequence, the electronic distributions around the Fermi levels of the two ESs gradually smear out (or, physically, heat up) as outlined in Fig.~\ref{fig:1} and its caption.~Our analysis is relevant for a steady-state situation and focuses on weakly broken translational invariance along the edge. We provide a step-by-step understanding of the energy exchange processes, i.e. as a function of the distance between the QDs. This is even achieved analytically under certain simplifying assumptions. In fact, the analytical distribution functions are found in a non-perturbative and far-from-equilibrium situation.\cite{footnote-advantage-of-blob} Finally, we note that both the energy resolved injection and detection of electrons using side-coupled QDs have already been realized in separate experiments.\cite{Altimiras-PRL-2012,Tewari-PRB-2016,Kiyama-PRB-2015}  A system combining both capabilities -- very similar to our setup -- is being currently investigated.\cite{Tewari-PRB-2016} Hence our predictions could be tested in the near future.

\subsection{Main results and experimental predictions}

We analyze the relaxation between two ESs where electrons are injected into one of them high above the Fermi energy of an otherwise equilibrated system.  The relaxation dynamics is described by two coupled (nonlinear integro-differential) kinetic equations, which we reduce to a more intuitive set of coupled Fokker-Planck equations (Secs. \ref{sec:coupl-fokk-planck} and \ref{sec:coupl-fokk-planck-1}). This is possible basically by using two conditions: (i) The translational invariance is only weakly broken and (ii) the injection QD (see Fig.~\ref{fig:1}) is only weakly coupled to the ES. The weakly broken translational invariance means that the energy scale $\Delta E$ of the allowed energy exchange per collision (due to the non-momentum conserving scattering) becomes the smallest energy scale in the problem. The weak coupling between the QD and the outer ES is controlled by gate voltages and ensures that the EDIE is small compared to full occupation (i.e.~one). Thus, Pauli-blocking for the injected electrons can be neglected.\cite{footnote-non-linear-FP-eq}

From our analysis,  the physical picture illustrated in Fig.~\ref{fig:1} emerges: The  EDIE with average energy $E_0$ in the outer ES relaxes via a generalized drift-diffusion process. The energy lost by the injected electrons smears out the distribution of the inner ES around the Fermi level $\mu_i$, i.e.~intuitively heats up the inner ES [Fig.~\ref{fig:1}(a-b)]. The heat of the inner ES in turn gradually equilibrates with the outer ES, which therefore also smears out around its (initial) Fermi-level $\mu_o$ [Fig.~\ref{fig:1}(c-d)].\cite{footnote-no-intra-ES-scattering} The supplementary material contains movies showing the described course of the relaxation (See e.g. \texttt{symmetric.avi}).\cite{supp-mat}  

A simple prediction of our work, which should be easy to verify experimentally, is that the average energy $\langle E\rangle$ of the EDIE moves towards the Fermi-level as [see Fig.~\ref{fig:1}(d)]
\begin{equation}
\label{eq:averge-E}
\langle E\rangle=E_0-\eta x,
\end{equation}
i.e.~with \emph{constant drift-velocity} in energy space given by
\begin{equation}
\label{eq:3}
\eta = \frac{\sqrt{\pi}}{4}\gamma(\Delta E)^3,
\end{equation}
where $\gamma$ is the effective inter-ES interaction (specified later in Eq.~(\ref{eq:def:gamma-and-DE})) and $x$ is the distance between the QDs. The average energy is equal to the maximum of the EDIE, $E_{\textrm{max}}(x)=\langle E\rangle$, if it is even around its maximum initially, which is a very likely experimental\cite{Tewari-PRB-2016} realization.  

Furthermore, while the EDIE loses energy on average, its width increases as
\begin{align}
\label{eq:4}
\Gamma(x) = \sqrt{\Gamma_0^2+4\eta\int_{0}^xdx'D_{i}(0,x')},
\end{align}
where $\Gamma_0$ is the initial width of the EDIE and $D_i(0,x)$ describes the energy smearing around the Fermi level of the inner ES, i.e.~intuitively the temperature of the inner ES (defined rigorously in Sec.~\ref{sec:model}). The function $D_i(0,x)$ is nothing else than the dynamical diffusion parameter of the drift-diffusion relaxation process in the inner ES. Interestingly, the spreading of the EDIE is \emph{dynamically accelerated} during the relaxation,  since the heating of the inner ES [i.e.~increasing $D_i(0,x)$] is caused by the gradual energy loss of the injected electrons. Due to this feedback mechanism, we dub the drift-diffusion process of the EDIE to be generalized compared to e.g. Brownian motion.\cite{Risken-BOOK} Strictly speaking, Eq.~(\ref{eq:4}) \emph{only} holds for an initial Gaussian EDIE, but other forms of the distribution give qualitatively the same result (see Appendix \ref{app:initial-blobs}).
 
We deal with the relaxation and thereby the smearing of the inner ES, $D_i(0,x)$, using two different models. In the first model in Sec.~\ref{sec:coupl-fokk-planck}, we neglect the Fermi sea of the outer ES and thereby only focus on the interplay between the injected electrons and the inner ES. The advantage of this model is \emph{analytic} distribution functions for \emph{both} ESs, showing e.g.~that the inner ES strictly speaking does \emph{not} remain an exact Fermi distribution with elevated temperature for $x>0$ --- even though it is close.  The problem of this model is that the inner ES smears out too fast (i.e.~over-heats), which acts back on the EDIE by inducing too fast an energy spreading via $D_i(0,x)$, see Eq.~(\ref{eq:4}). The over-heating problem is taken care of in our second model in Sec.~\ref{sec:coupl-fokk-planck-1}, where the Fermi sea of the outer ES is included again and therefore is able to receive some of the energy from the smearing of the Fermi sea of the inner ES. We emphasize that Eqs.~(\ref{eq:averge-E}-\ref{eq:4}) are valid for both models, and only the widening of the EDIE differs due to different $D_i(0,x)$. However, the second model only offers immediate analytic solution of the EDIE, but not for the entire electronic distribution of both ESs. Nevertheless, by introducing an \emph{effective temperature approach} in Sec.~\ref{sec:effect-temp-appr}, we gain powerful physical insight into the energy redistribution between the ESs and the electronic distributions.  Finally, we compare this approach with an exact iterative numerical solution of the original kinetic equations.

Our Fokker-Planck approach is valid as long as the EDIE is sufficiently well separated in energy from  the Fermi sea of the outer ES. Once the EDIE gets close to the Fermi sea, then the relaxation scheme discussed above changes. This happens at a characteristic length 
\begin{align} 
\ell_{\textsc{fp}}=(E_0-\mu_o)/\eta,   
\end{align} 
where the average energy of the injected electrons equals the Fermi level. We emphasize that  
$\ell_{\textsc{fp}}$ characterizes nearly full relaxation for our setup with an EDIE and depends explicitly on the initial energy $E_0$, i.e. it is \emph{not} a generic relaxation length scale in ESs. In contrast, the velocity in energy space, $\eta$, is a more fundamental quantity obtained within our theory.

We treat the relaxation of the EDIE close to the Fermi level, $x\gtrsim\ell_{\textsc{fp}}$, by noting that two Fermi distributions with equal temperatures solve the full kinetic equations. The common temperature [see Eq.~(\ref{eq:T-infty})] and the Fermi levels of the fully equilibrated system are found from conservation of energy and particle number (Sec.~\ref{sec:full-relax}). We confirm that the Fermi distributions indeed are the equilibrated distributions by numerically solving the full coupled kinetic equations (Sec.~\ref{sec:effect-temp-appr}). Finally, we discuss relaxation beyond the Fokker-Planck regime and electron-hole symmetric relaxation in Sec.~\ref{sec:discussion}. 

\section{The model}\label{sec:model}

To analyze the electronic relaxation within the ESs, we use the Boltzmann kinetic equation
\begin{align}
\label{Boltzmann-eq}
v_{k\alpha}\p_x f_{\alpha}(k,x)=I_{kx\alpha}[f_\alpha,f_{\bar\alpha}]
\end{align}
where $f_\alpha$ is the distribution function for the inner ($\alpha=i$) or outer ($\alpha=o$) ES at position $x$ and wavenumber $k$. The velocity is $v_{k\alpha}=(1/\h)\p_kE_{k\alpha}$ and $\bar\alpha$ is the opposite ES of $\alpha$. The energy exchange between the ESs is mediated by two-particle inter-ES interactions described by the collision integral\cite{footnote-index-specification} 
\begin{align}
I_{k_1^{}x\alpha}[f_\alpha,&f_{\bar\alpha}]=
\sum_{k_2^{}k_{1'}k_{2'}}
W_{1^{}2^{},{1'}{2'}}\\
&\times
\Big\{
f_{\alpha}(k_{1'},x)[1-f_{\alpha}(k_{1}^{},x)]
f_{\bar{\alpha}}(k_{2'},x)[1-f_{\bar\alpha}(k_{2}^{},x)]
\nonumber\\
&-
f_{\alpha}(k_{1}^{},x)[1-f_{\alpha}(k_{1'},x)]
f_{\bar{\alpha}}(k_{2},x)[1-f_{\bar\alpha}(k_{2'},x)]
\Big\}.
\nonumber
\end{align}
The Pauli-principle is incorporated into the combination of distribution functions by the in- and out-scattering terms as $f_{\alpha}(k,x)[1-f_{\alpha}(k',x)]$, which describes a scattering from $k$ to $k'$ in ES $\alpha$. The scattering rate $W_{1^{}2^{},{1'}{2'}}$ is found from the Fermi golden rule, i.e. $W_{1^{}2,{1'}{2'}}=\frac{2\pi}{\h}|\langle k_{1'}\alpha,k_{2'}\bar\alpha|V|k_{1}\alpha, k_{2}\bar\alpha\rangle|^2\delta(E_{k_{1}\alpha}+E_{k_{2}\bar\alpha}-E_{k_{1'}\alpha}-E_{k_{2'}\bar{\alpha}})$, from which the conservation of total energy in the scattering is explicit. 

As in our previous work,\cite{Lunde-2010} we incorporate the non-translational invariance by using an interaction between electrons in the inner and outer ESs at positions $x$ and $x'$, respectively, of the form $V(x,x')=V_0\delta(x-x')g(x)$.  Here $V_0$ is the interaction strength and $g(x)$ is a dimensionless function modeling the variations of the inter-ES interaction along the edge due to the lack of translational invariance, which opens the possibility of non-momentum conserving scattering.\cite{footnote-trans-inv-int}  We are not interested in studying a specific disorder realization, so we average over different configurations and assume a Gaussian distributed deviation of $g$ from its mean $g_0$, i.e. $\overline{(g(x)-g_0)(g(x')-g_0)}=A/(\sqrt{2\pi}\ell_p) e^{-(x-x')^2/(2\ell_p^2)}$. Here we introduce the momentum-breaking correlation length $\ell_p$ characterizing the amount of non-momentum conservation. The length $A$ describes the typical magnitude of the interaction variations. This leads to an averaged interaction matrix element with a non-momentum conserving part of the form  $\overline{|\langle k_{1'}\alpha,k_{2'}\bar\alpha|V|k_{1}^{}\alpha, k_{2}^{}\bar\alpha\rangle|^2}_{\Delta k\neq0} \propto V_0^2 A e^{-\Delta k^2\ell_p^2/2}$, where $\Delta k$ is the difference between the total momentum before and after the scattering, $\Delta k\equiv k_{1}^{}+k_{2}^{}-k_{1'}-k_{2'}$. Therefore, the larger $\ell_p$ is, the more restricted is the possibility for having $\Delta k\neq0$. This interaction model is convenient, but its specific form is not important for the physics described in this paper as long as it incorporates the non-momentum conserving processes. Finally, we note that the \emph{intra}-ES interaction matrix element is zero due to cancellation of the direct and exchange interaction terms. This cancellation depends somewhat on the interaction model, but the intra-ES interaction is generally strongly suppressed. Further details on the used interaction are found in Ref.~\onlinecite{Lunde-2010} and its supplementary material.\cite{supp-mat}   

The phase space for momentum \emph{and} energy conserving scattering is very limited in one dimension, so the functional form of the dispersion relation $E_{k\alpha}$ plays a role for the detailed relaxation.\cite{Lunde-PRB-2007,Karzig-Levchenko-Glazman-Oppen-NJP-2012} However, here the phase space opens up due to the non-momentum conserving processes and the form of the dispersion relation is less important. Thus, we simply use linear dispersions $E_{k\alpha}=\mu_\alpha+\h v_\alpha (k-k_{F,\alpha})$,\cite{footnote-confinement-pot} and from now on consider the kinetic equation (\ref{Boltzmann-eq}) in energy space, i.e. 
\begin{subequations}
\begin{align}
\label{eq:1}
\partial_x f_{\alpha}(E,x)=\mathcal{I}_{Ex\alpha}[f_\alpha,f_{\bar\alpha}].
\end{align}
Here  $1/v_\alpha$ is absorbed into the collision integral as indicated by the notation change, $I\rightarrow\mathcal{I}$, i.e.
\begin{align}
\label{eq:2}
\mathcal{I}_{Ex\alpha}[f_\alpha,f_{\bar\alpha}]
=&
\gamma\int_{-\infty}^{\infty}d\om\,
e^{-(\om/\Delta E)^2}
\\
&\times
\Big\{f_{\alpha}(E+\om,x)[1-f_{\alpha}(E,x)]D_{\bar\alpha}(\omega,x)
\nonumber\\
& \ \ \ 
-f_{\alpha}(E,x)[1-f_{\alpha}(E+\omega,x)]D_{\bar\alpha}(-\omega,x)\Big\},
\nonumber
\end{align}
\label{full-KE}%   %HERE THE "%" RIGHT AFTER THE LABEL MUST NOT BE REMOVED!!! 
\end{subequations}
which describes the exchange of an energy $\om$ between the two ESs. The available phase space in ES $\bar\alpha$ to absorb an energy $\om$ is accounted for by
\begin{equation}
\label{eq:20}
D_{\bar\alpha}(\omega,x)=
\int_{-\infty}^{\infty}dE'\; f_{\bar\alpha}(E'-\omega,x)[1-f_{\bar\alpha}(E',x)].
\end{equation}
Similarly, $D_{\bar\alpha}(-\omega,x)$ gives the phase space for emitting energy $+\om$ by ES $\bar\alpha$. Moreover, 
\begin{equation}
\label{eq:def:gamma-and-DE}
\gamma =\frac{V_0^2A}{(2\pi\h v_i\h v_o)^2}
\quad \textrm{and}\quad
\Delta E=\frac{\sqrt{2}}{\ell_p}\frac{\h v_iv_o}{|v_i-v_o|}
\end{equation}
are, respectively, the effective interaction strength and the energy scale for the allowed energy exchange per collision due to the non-momentum conserving scattering.

Mathematically, our problem is now to solve the two coupled kinetic Eqs.~(\ref{full-KE}) for the distribution functions $f_\alpha(E,x)$ of the inner and outer ESs ($\alpha=i,o$) with the initial condition that $f_i(E,0)$ is a Fermi distribution and $f_o(E,0)$ is a Fermi distribution plus the initial EDIE $b(E,0)$. In general, this is a hard problem to solve analytically. Our Fokker-Planck approach however offers interesting insight into the relaxation process.

\subsection{Conserved quantities}\label{sec:conserved-quan}

The kinetic equation~(\ref{Boltzmann-eq}) leads to conservation of (i) the particle number of each ES separately, and (ii) the total energy of both ESs. These conserved quantities are very useful guidelines to both understand the relaxation and make meaningful approximations. Therefore, we briefly discuss them here. 

Conservation of the number of particles in a single ES, $N_\alpha= \int dE \rho_\alpha f_{\alpha}(E,x)$, where $\rho_\alpha=L/(2\pi\h v_\alpha)$ is the density of states, follows from
\begin{align} 
\label{particle-conservation-general}
\p_x\int_{-\infty}^{\infty} dE f_{\alpha}(E,x)=0.
\end{align}
This is shown by using the kinetic equation (\ref{Boltzmann-eq}) in Appendix \ref{appendix:conserved}. Due to the linear dispersions, it can be convenient to introduce a low-energy cut-off such that $N_\alpha$ is finite. This does not affect the physics discussed here.  

Even though the ESs exchange energy, the sum of the electronic energies in the ESs is conserved in each collision, i.e. $E_{k_{1}^{}\alpha}+E_{k_{2}^{}\bar\alpha}=E_{k_{1'}\alpha}+E_{k_{2'}\bar{\alpha}}$. Therefore, the total energy current is conserved, which leads to the following conserved quantity (derived in Appendix \ref{appendix:conserved}): 
\begin{subequations}
\begin{align} 
\p_x \sum_{\alpha} Z_{\alpha}(x)&=0, 
\qquad \textrm{where} \qquad\\
Z_{\alpha}(x)&=
\int_{-\infty}^{\infty} d E (E-\bar\mu_\alpha)f_\alpha(E,x). 
\end{align}
\label{energy-conservation-general}%   %HERE THE "%" RIGHT AFTER THE LABEL MUST NOT BE REMOVED!!!
\end{subequations}
Here we introduced an (arbitrary) reference chemical potential $\bar\mu_\alpha$ for calculational convenience. We formulated the conservation laws in Eqs.~(\ref{particle-conservation-general}-\ref{energy-conservation-general}) for our specific case. However, in Appendix \ref{appendix:conserved}, they are derived from the kinetic equation (\ref{Boltzmann-eq}) with general dispersion relations.

\section{Relaxation of the injected electrons without a Fermi sea in the outer edge state\label{sec:coupl-fokk-planck}}
  
In this section, we consider a simplified problem, where we ignore the
   Fermi sea in the outer ES, into which the electrons are injected. This simplification leads to analytic solutions for the distribution functions. It includes the physics illustrated in Fig.~\ref{fig:1}(a) and (b), but excludes the energy-redistribution between the Fermi seas shown in Fig.~\ref{fig:1}(c) and (d).

\subsection{The coupled Fokker-Planck kinetic equations\label{sec:coupl-fokk-planck-model1-derivation}}

\begin{figure*}
\includegraphics[width=0.8\columnwidth]{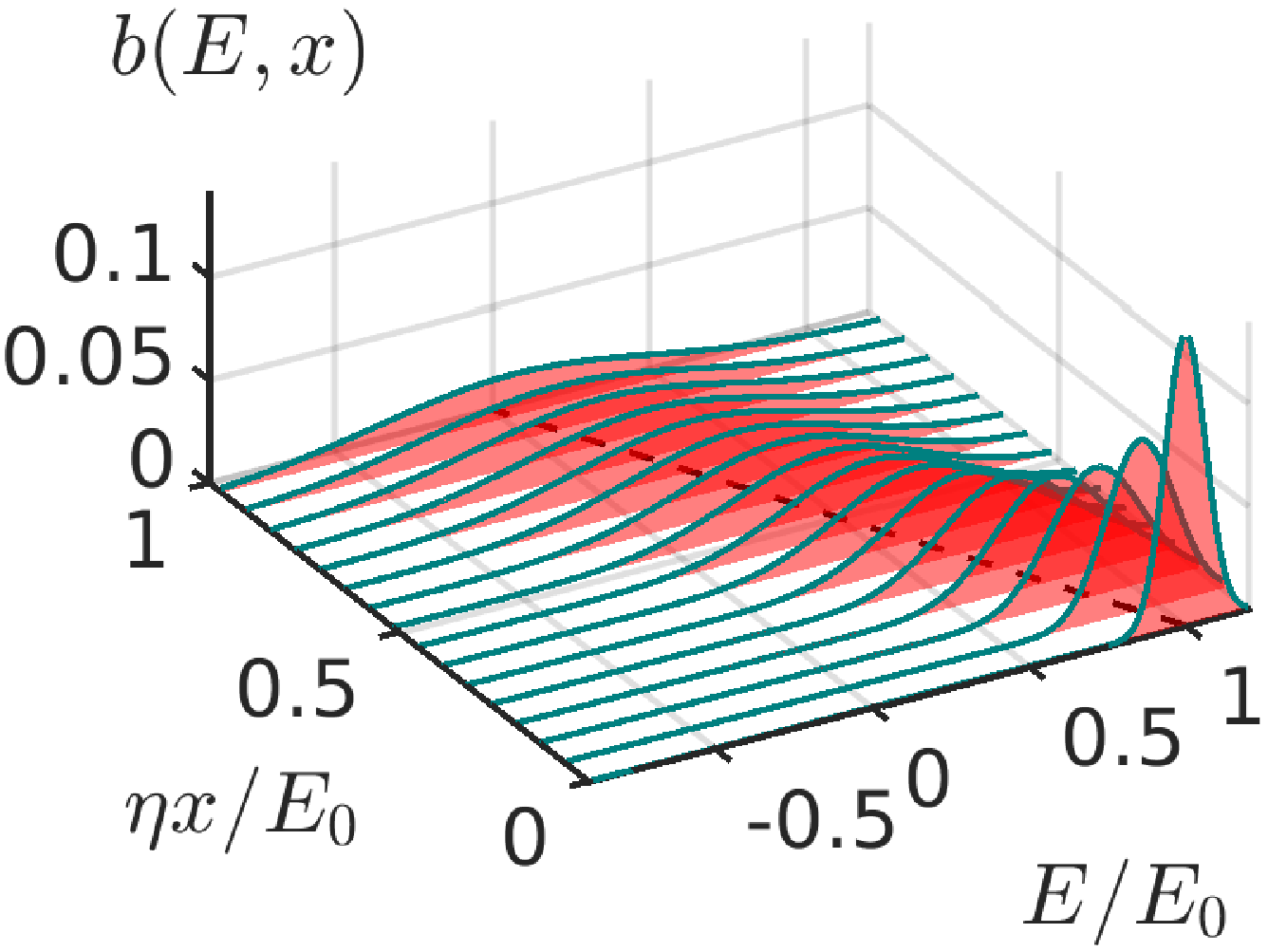}\hspace{2cm}
\includegraphics[width=0.8\columnwidth]{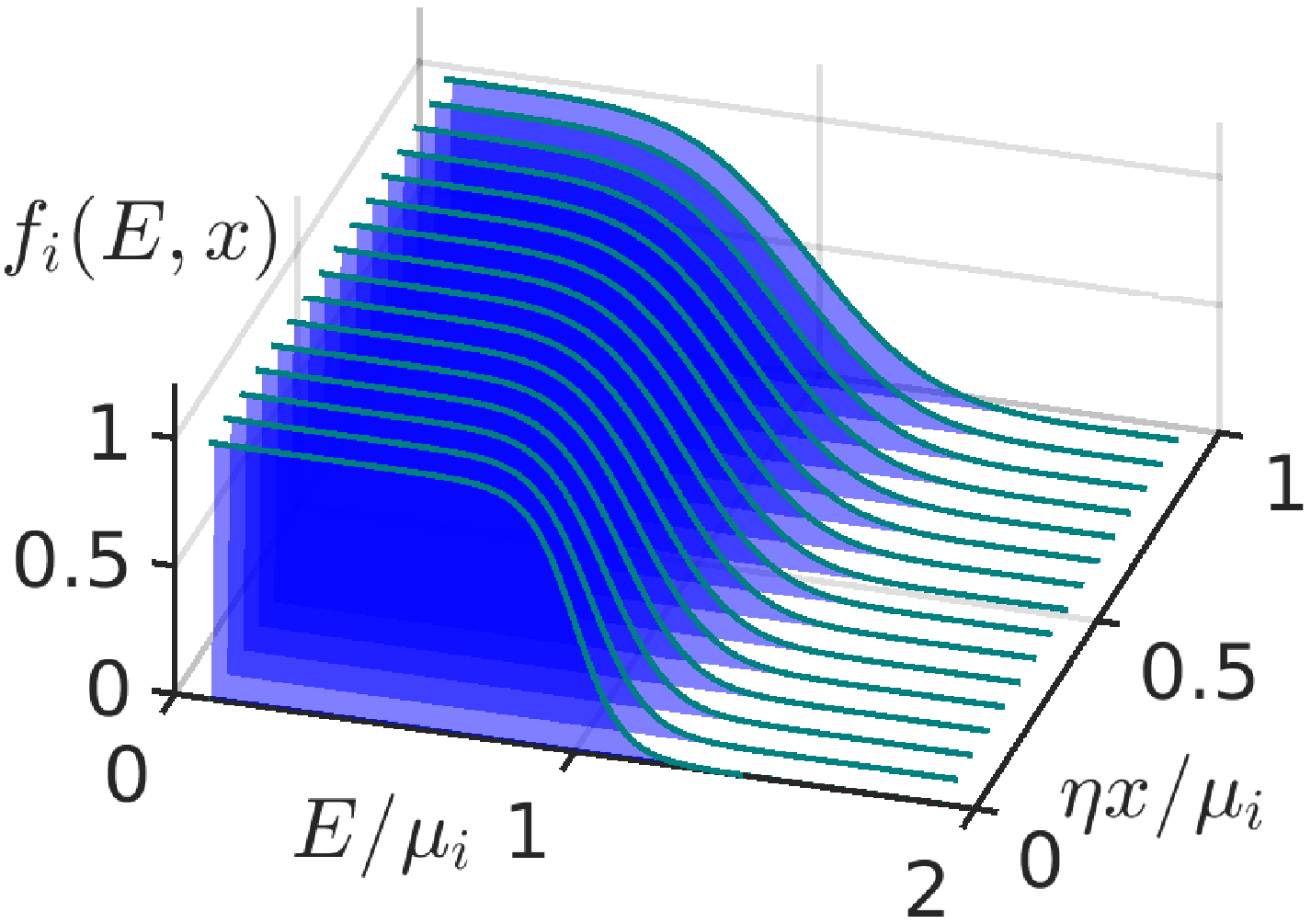}
\caption{(Color online)
Analytic solution of the distribution functions for the model without a Fermi sea in the outer ES in Sec.~\ref{sec:coupl-fokk-planck}. Left panel: Evolution of the gaussian EDIE $b(E,x)$ in Eq.~(\ref{eq:15}) with the width given in Eq.~(\ref{eq:4}) and the energy smearing of the inner Fermi sea $D_i(0,x)$ in Eq.~(\ref{eq:28}). The dashed (black) line indicates the evolution of the maximum of the distribution given by Eq.~(\ref{eq:max-gauss}). Right panel: corresponding evolution of the inner ES distribution $f_i(E,x)$ Eq.~(\ref{eq:29}).  As long as $\Delta E$ is the smallest energy scale of the problem, the distribution functions only depend on $\eta$ (and thereby $\Delta E$ and $\gamma$) through the unit of $x$. Thus, using $E_0=\mu_i$ in the figure, we can scale $x$ by the same unit of length, namely $E_0/\eta=\mu_i/\eta$. The parameters are (in arbitrary units): $N_b/\rho_o=0.5$, $\Gamma_0=2$ and $\kb T=1.25$.
\label{fig:3}}
\end{figure*}

Within this model, the distribution function of the outer ES is restricted to only the EDIE (i.e.~a \emph{b}lob) and written as
\begin{equation}
\label{eq:8}
f_o(E,x)=b(E,x), 
\end{equation}
while the inner ES distribution function $f_i(E,x)$ initially is a Fermi function with temperature $T$. The particle number $N_b$ of the EDIE is conserved, so
\begin{equation}
\label{eq:34}
\int_{-\infty}^{\infty} d E\ b(E,x) 
=\frac{N_b}{\rho_o} 
\end{equation}
is independent of $x$. The coupled kinetic equations (\ref{full-KE})
\begin{subequations}
\begin{align}
\p_x f^{}_{i}(E,x)&= 
\mathcal{I}_{Exi}[f_i,b],
\label{dfi-KE-model1}\\
%%%%
\p_x b(E,x)&= 
\mathcal{I}_{Exo}[b,f_i],
\end{align}
\label{full-KE-model1}%   %HERE THE "%" RIGHT AFTER THE LABEL MUST NOT BE REMOVED!!! 
\end{subequations}
greatly simplify by assuming that the EDIE is small in the sense that $b(E,x)\ll 1$, such that Pauli-blocking can be neglected in the collision integrals: $b(1-b)\simeq b$. The smallness of the initial distribution $b(E,0)$ is controlled experimentally by the confinement gates of the injection QD. The collision integrals simplify to 
\begin{subequations}
\begin{equation}
\label{eq:23}
\mathcal{I}_{Exi}[f_i,b]
\simeq
\gamma \frac{N_b}{\rho_o} \!
\int_{-\infty}^{\infty}\!\!\! d  \om \ e^{-\left[\frac{\omega}{\Delta E}\right]^2}
\Big\{f_{i}(E+\omega,x)-f_{i}(E,x)\Big\},
\end{equation}
since $D_o(\omega,x)\simeq N_b/\rho_o$ for $b(E,x)\ll 1$,  and 
\begin{align}
\mathcal{I}_{Exo}[b,f_i]
\simeq  
\gamma &
\int_{-\infty}^{\infty} d  \omega \ e^{-\left[\frac{\omega}{\Delta E}\right]^2}
\label{eq:24}\\
&
\times
\Big\{b(E+\omega,x)D_{i}(\omega,x)-b(E,x)D_{i}(-\omega,x)\Big\}\nonumber.
\end{align}
\label{eq:coll-int-model1}%    %HERE THE "%" RIGHT AFTER THE LABEL MUST NOT BE REMOVED!!!
\end{subequations}
Next we restrict ourself to the limit of \emph{weakly broken translational invariance} such that $\Delta E$ becomes the smallest energy scale, in particular smaller than variations in energy of the EDIE $b(E,x)$, the distribution $f_i(E,x)$ as well as the function $D_{i}(E,x)$. Thus, the Gaussian kernel in (\ref{eq:coll-int-model1}) becomes a strongly peaked function around $\om=0$, such that only the lowest order terms in the energy exchange $\om$ in the curly brackets are needed. Formally, a new integration variable $z=\frac{\om}{\Delta E}$ can be introduced, which allows for an expansion in the small parameters given by $\Delta E$ over the energy variations of the involved functions. This is the idea behind deriving a Fokker-Planck equation from a rate-like equation.\cite{Breuer-Petruccione-BOOK,Micklitz-Jerome-Matveev-PRB-2010} After expanding and doing the integrals, the kinetic Eqs.~(\ref{full-KE-model1}) become
\begin{subequations}
\begin{align}
\p_x f^{}_{i}(E,x)&= 
\eta \frac{N_b}{\rho_o} \p^2_Ef_{i}(E,x),
\label{eq:25}\\
\p_x b(E,x)&=
\label{eq:26}
2\eta\Big[\p_{\omega}D_{i}(0,x)\p_Eb(E,x)+\frac{1}{2}D_{i}(0,x)\p_E^2b(E,x)\Big],
\end{align}
\label{eq:FP-model1}%HERE THE "%" RIGHT AFTER THE LABEL MUST NOT BE REMOVED!!!
\end{subequations}
where $\eta$ is defined in Eq.~(\ref{eq:3}). Using only that $f_i(E,x)$ is fully occupied (empty) for very low (high) energy, we get $\p_{\omega} D_i(0,x)=\frac{1}{2}$ independently of $x$ (see Appendix \ref{app:dD}). This reduces the kinetic equation (\ref{eq:26}) for $b(E,x)$ to 
\begin{align}
\p_x b(E,x)
= \eta
\Big\{\p_Eb(E,x)+D_{i}(0,x)\p_E^2b(E,x)\Big\},
\label{eq:30}
\end{align}
which is a Fokker-Planck equation.\cite{Risken-BOOK} Together with the kinetic equation (\ref{eq:25}) for the inner ES distribution $f_i$, these two simplified kinetic equations determine the evolution of the distributions in the small $\Delta E$ limit. In Eq.~(\ref{eq:30}), the first term in the brackets on the right-hand side describes the drift motion of the EDIE in energy with velocity $\eta$, while the second term describes its diffusion in energy with dynamical diffusion parameter $\eta D_{i}(0,x)$. The diffusion in energy is termed dynamical due to its $x$-dependence. The kinetic equations (\ref{eq:25}) and (\ref{eq:30}) still couple, but in a much weaker way than the original kinetic equations  (\ref{full-KE}) with the full collision integrals. In fact, the evolution of the inner distribution $f_i$ in Eq.~(\ref{eq:25}) \emph{only} couples to the EDIE via its particle number $N_b$ --- not to the full form of $b(E,x)$ as in the original kinetic equation. This stems from the smallness of the EDIE and already appears in the collision integral approximation (\ref{eq:23}).  The evolution of the EDIE in Eq.~(\ref{eq:30}) couples to the other ES via $D_{i}(0,x)$, which describes the smearing of the distribution of the inner ES around the Fermi level and initially $D_{i}(0,0)=\kb T$.  

\subsection{The analytic distribution functions}

Now we discuss the solutions of the coupled Fokker-Planck equations (\ref{eq:25}) and (\ref{eq:30}). Eq.~(\ref{eq:25}), which is a one-dimensional diffusion equation, is readily solved to yield
\begin{equation}
\label{eq:29}
f_i(E,x)=\sqrt{\frac{\rho_o}{4\pi N_b \eta x}}
\int^{\infty}_{-\infty}  dE'\ e^{-\frac{(E-E')^2}{4 N_b\eta x/\rho_o}}
f_{i}(E',0),
\end{equation}
where $f_{i}(E',0)$ is the initial Fermi distribution of temperature $T$. Interestingly, this solution explicitly shows that $f_i(E,x)$ \emph{does not remain a Fermi distribution with elevated temperature for} $x>0$, even though it is rather similar as seen on Fig.~\ref{fig:3}. Nevertheless, once the injected electrons fully relax at $x>\ell_{\textsc{fp}}$, then $f_i(E,x)$ becomes a Fermi distribution again, see Sec.~\ref{sec:effect-temp-appr}. However, full relaxation is \emph{never} reached within the present model, since it is the missing Fermi sea in the outer ES that effectively ends the relaxation.   

The Fokker-Planck Eq.~(\ref{eq:30}) for the EDIE can be solved analytically, \emph{if} we assume an initially Gaussian form of $b(E,0)$. In that case, the EDIE remains Gaussian for $x>0$, 
\begin{equation}\label{eq:15}
b(E,x)=\frac{N_b}{\Gamma(x)\rho_o\sqrt{\pi}}
\exp\!\left[-\frac{[E-E_0(x)]^2}{\Gamma(x)^2}\right].
\end{equation}
To obtain $E_0(x)$ and $\Gamma(x)$, we insert $b(E,x)$ in Eq.~(\ref{eq:15}) into Eq.~(\ref{eq:30}) and solve the two simple differential equations that arise. We find the maximum of the EDIE $E_0(x)$ to gradually move downwards in energy, 
\begin{equation}
\label{eq:max-gauss}
E_0(x)=-\eta x+E_0,
\end{equation} 
from its initial value $E_0$, while the width  $\Gamma(x)$ increases according to Eq.~(\ref{eq:4}). The constant drift-velocity $\eta$, given in Eq.~(\ref{eq:3}), only depends on the effective inter-ES interaction and  $\Delta E$, while the width of the distribution $\Gamma(x)$ also depends on the inner ES distribution through its smearing $D_i(0,x)$. By inserting $f_i(E,x)$ from Eq.~(\ref{eq:29}) into the definition of $D_i(0,x)$ in Eq.~(\ref{eq:20}), we find
\begin{align}
\label{eq:28}
&D_{i}(0,x)
=
\sqrt{\frac{\rho_o}{8\pi N_b\eta x}}
\int_{-\infty}^{\infty} d E\
e^{-\frac{\rho_o E^2}{8 N_b\eta x}}
\frac{E}{1-e^{-E/\kb T}}.
\end{align}
This shows that the energy smearing of the inner ESs Fermi level initially is nothing but the temperature, $D_{i}(0,0)=\kb T$, and that it increases for longer distances, $D_{i}(0,x)>D_{i}(0,0)$. In the limit of long distances, $\sqrt{8N_b\eta x/\rho_o}\gg \kb T$, we approximately find $D_i(0,x)\approx\sqrt{2N_b\eta x/(\pi\rho_o)}$ (keeping $\Delta E\ll \kb T$ in mind). Thereby, we have a complete description of the evolution of the EDIE $b(E,x)$ and the inner ES distribution $f_i(E,x)$ within this simplified model as shown in Fig.~\ref{fig:3}. 

\subsection{Generic features of the relaxation}

The relaxation pattern of the EDIE does not depend crucially on the assumption of an initially Gaussian distribution. An initial Lorentzian distribution or an initial distribution $\propto\!\cosh^{-2}[(E-E_0)/(2\Gamma_{\rm c})]$ do not change our predictions qualitatively as discussed in Appendix \ref{app:initial-blobs}. Furthermore, for \emph{any} initial EDIE, the average energy of the injected electrons, 
\begin{align} 
\langle E\rangle\equiv 
\frac{\rho_o}{N_b} 
\int_{-\infty}^{\infty}d E \; E b(E,x), 
\label{eq:def-average-E}
\end{align}
is found to evolve in space like the maximum of the Gaussian distribution in Eq.~(\ref{eq:max-gauss}), i.e.~$\langle E\rangle=-\eta x+E_0$. This can be realized by inserting Eq.~(\ref{eq:30}) into $\p_x\langle E\rangle$, i.e. 
\begin{align} 
\p_x\langle E\rangle
&=
\frac{\rho_o}{N_b} \int_{-\infty}^{\infty}
\!\!\!\!d E E 
\eta
\Big\{\p_Eb(E,x)+D_{\bar{\alpha}}(0,x)\p_E^2b(E,x) \Big\}
=-\eta, 
\label{eq:average-E-evolution}
\end{align}
where we used partial integration and the fact that $b(E,x)\rightarrow0$, $Eb(E,x)\rightarrow0$ and $E\p_Eb(E,x)\rightarrow0$ for $E\rightarrow\pm\infty$. We emphasize that the evolution of the average energy $\langle E\rangle$ is generic for any EDIE. If we specialize to an EDIE that is initially symmetric in energy around its maximum $E_{\textrm{max}}(x)$, then we find that $\langle E\rangle=E_{\textrm{max}}(x)$ and that an initially symmetric distribution remains symmetric for $x>0$. This is shown in Appendix \ref{app:symmetric-packet}.      

Finally, we observe that energy is transferred by the relaxation of the injected electrons to the inner ES at a constant rate, which is reflected by the fact that 
\begin{align}
\p_xZ_o(x)=-\p_xZ_i(x)=-\eta \frac{N_b}{\rho_o}.
\label{eq:dZ-model1}
\end{align}
This is shown by inserting the Fokker-Planck Eqs.~(\ref{eq:25}) and (\ref{eq:30}) into the definition of $Z_\alpha(x)$ in Eq.~(\ref{energy-conservation-general}) (similar derivation as given Appendix \ref{app:dZ}). This gradual energy transfer, which causes smearing of the distribution around the inner ES's Fermi level,   is physically interpreted as heating of the inner ES. Moreover, Eq.~(\ref{eq:dZ-model1}) also shows that the conservation law $\p_x[Z_o(x)+Z_i(x)]=0$ in Eq.~(\ref{energy-conservation-general}) is fulfilled irrespectively of the form of $b(E,0)$ as expected.  

\section{Relaxation of the injected electrons including the Fermi sea in both edge states\label{sec:coupl-fokk-planck-1}}

Next, we incorporate the outer ES's Fermi sea and thereby analyze the relaxation of the injected electrons in the presence of Fermi seas in both ESs. Although we consider a EDIE with an average energy high above the Fermi level,  the Fermi sea of the outer ES still plays a role, since it absorbs energy from the inner ES. The physical picture emerging from our analysis is that the relaxation of the injected electrons heats the inner ES and subsequently the heat redistributes between the two ESs as shown in Fig.~\ref{fig:1}. Here, heating refers to an increase of the energy smearing around the Fermi level. The reduced heating of the inner ES causes the EDIE to spread less than described by the model without the outer Fermi sea in Sec.~\ref{sec:coupl-fokk-planck}.

\subsection{The coupled Fokker-Planck kinetic equations}\label{subsec:FP-model2}

We begin by separating the outer ES distribution $f_o$ into an EDIE $b(E,x)$ and a function $\fo(E,x)$ describing the Fermi sea:  
\begin{align}
\label{eq:5}
f_{o}(E,x)=\fo(E,x)+b(E,x),  
\end{align}
where $\fo$ as well as $f_i$ initially are Fermi functions of temperature $T$ and chemical potentials $\mu_{o}$ and $\mu_i$, i.e. 
\begin{align}
\label{eq:9}
\fo(E,x=0)
= \frac{1}{1+\exp\left(\frac{E-\mu_{o}}{\kb T}\right)}. 
\end{align}
The Fermi energies of the two ESs can easily be tuned separately in an experiment, although they are equal for the cartoon setup in Fig.~\ref{fig:1}(e). Using the rewriting (\ref{eq:5}), the full coupled kinetic equations (\ref{full-KE}) become
\begin{subequations}
\begin{align}
\partial_x f_{i}(E,x)
&=
\mathcal{I}_{Exi}[f_i,\fo+b],
\\
\partial_x [\fo(E,x)+b(E,x)]
&=
\mathcal{I}_{Exo}[\fo+b,f_i],
\label{eq:outer-KE}
\end{align}
\label{full-KE-model2}% DO NOT REMOVE %
\end{subequations}
which describe the entire relaxation until equilibrium.

Here, we specialize to the case of an EDIE high above the Fermi level and  sufficiently narrow such that $b(E,x)$ and $\fo(E,x)$ do \emph{not} overlap in energy. As the EDIE loses energy on average and widens, this assumption will eventually break down, so the final stages of the relaxation are excluded from the description below. The separability in energy of $b$ and $\fo$ is taken to be much larger than $\Delta E$ such that the particle number in each distribution is  conserved separately, i.e.
\begin{align}
\int_{-\infty}^{\infty} d E\ b(E,x)&=\frac{N_b}{\rho_o}
\quad \textrm{and}\quad 
\int_{\epsilon_c}^{\infty}d E\ \fo(E,x)=\frac{N_{\textsc{fs}}}{\rho_o} 
\end{align}
are both independent of $x$. (Here $\epsilon_c\ll \mu_o,\mu_i$ is a low energy cutoff to keep $N_{\textsc{fs}}<\infty$ and without importance for the described physics.)  The separation in energy between $b$ and $\fo$ allows us to neglect products of the kind $b(E\pm\om,x)\fo(E,x)$ in the collision integrals in Eq.~(\ref{full-KE-model2}) such that 
\begin{subequations}
\begin{align}
\mathcal{I}_{Exi}[f_i,\fo+b],
&\simeq
\mathcal{I}_{Exi}[f_i,\fo]+\mathcal{I}_{Exi}[f_i,b],
\\
\mathcal{I}_{Exo}[\fo+b,f_i]
&\simeq
\mathcal{I}_{Exo}[\fo,f_i] + \mathcal{I}_{Exo}[b,f_i].
\end{align}
\end{subequations}
Moreover, $\p_x \fo(E,x)$ and $\p_xb(E,x)$ naturally also separate in energy such that we can split the kinetic equation (\ref{eq:outer-KE}) into a kinetic equation for each of the two distributions $b$ and $\fo$, i.e. 
\begin{subequations}
\begin{align}
\partial_x f_{i}(E,x)
&=
\mathcal{I}_{Exi}[f_i,\fo]+\mathcal{I}_{Exi}[f_i,b],
\label{eq:dfi-model2}
\\
\partial_x \fo(E,x)
&=
\mathcal{I}_{Exo}[\fo,f_i],
\label{eq:dfo-model2}\\
\p_x b(E,x)
&=
\mathcal{I}_{Exo}[b,f_i]. 
\label{eq:db-model2}
\end{align}
\label{KE-model2}% DO NOT REMOVE %
\end{subequations}
Thus, the injected electrons and the outer ES's Fermi sea do not exchange energy directly, since no collision integral connect $b$ and $\fo$ and intra-ES interaction is absent (see Sec.~\ref{sec:model}). However, \emph{indirectly} they exchange energy through the inner ES. The appearance of $\fo$ and the ability of the Fermi seas of the two ESs to exchange energy due to $\mathcal{I}_{Exi}[f_i,\fo]$ and $\mathcal{I}_{Exo}[\fo,f_i]$ in Eq.~(\ref{KE-model2}) are new ingredients of this model [compare to  Eq.~(\ref{full-KE-model1})]. Furthermore, $\mathcal{I}_{Exo}[\fo,f_i]=\mathcal{I}_{Exi}[f_i,\fo]=0$ at $x=0$, since the collision integral of equal temperature Fermi functions vanish (even for $\mu_i\neq\mu_o$). Thus, the initial evolution at $x=0$ of the EDIE and $f_i$ is the same as for the model in Sec.~\ref{sec:coupl-fokk-planck}, since Eqs.~(\ref{eq:dfi-model2}) and (\ref{eq:db-model2}) simplify to Eqs.~(\ref{full-KE-model1}) at $x=0$.  This shows that some initial relaxation of the injected electrons is necessary to change $f_i$, before the Fermi seas begin to exchange energy. 

As in Sec.~\ref{sec:coupl-fokk-planck}, we now concentrate on the case of a small EDIE, $b(E,x)\ll1$, and therefore neglect Pauli-blocking, $b(1-b)\simeq b$, such that the collision integrals $\mathcal{I}_{Exi}[f_i,b]$ and $\mathcal{I}_{Exo}[b,f_i]$ simplify to the forms in Eq.~(\ref{eq:coll-int-model1}). Next, we consider the limit of weakly broken translational symmetry, such that $\Delta E$ is the smallest energy scale of the problem. By the same strategy as discussed after Eq.~(\ref{eq:coll-int-model1}), we  now derive a set of coupled Fokker-Planck equations from the kinetic Eqs.~(\ref{KE-model2}). The small $\Delta E$ limit of the collision integrals involving $b$ are the same as found previously in Eqs.~(\ref{eq:FP-model1}) and (\ref{eq:30}), i.e.   
\begin{subequations}
\begin{align}
\mathcal{I}_{Exi}[f_i,b]
&\simeq 
\eta \frac{N_b}{\rho_o} \p^2_Ef_{i}(E,x),
\\
\mathcal{I}_{Exo}[b,f_i]
&\simeq
\eta\Big[\p_Eb(E,x)+D_{i}(0,x)\p_E^2b(E,x)\Big].
\end{align}
\label{eq:FP-coll-int-with-b-model2}%HERE THE "%" RIGHT AFTER THE LABEL MUST NOT BE REMOVED!!!
\end{subequations}
The collision integrals describing the relaxation between the two Fermi seas in the limit of small $\Delta E$ are
\begin{subequations}
\begin{align}
\mathcal{I}_{Exi}[f_i,\fo]
&\simeq
\eta
\Big\{[1\!-\!2f_{i}(E,x)]\p_Ef_{i}(E,x) 
+\frak{D}_{o}(0,x)\p_E^2f_{i}(E,x)\Big\},
\\
\mathcal{I}_{Exo}[\fo,f_i]
&\simeq
\eta
\Big\{[1\!-\!2\fo(E,x)]\p_E\fo(E,x)+D_{i}(0,x)\p_E^2\fo(E,x)\Big\},
\end{align}
\label{eq:coll-int-Fermi-seas}% DO NOT REMOVE %
\end{subequations}
where in analogy to $D_{\alpha}(\omega,x)$ in Eq.~(\ref{eq:20}) we introduced  
\begin{align}
\frak{D}_{o}(\omega,x)=\int_{-\infty}^{\infty}dE'\; \fo(E'-\omega,x)[1-\fo(E',x)]
\label{eq:def-Dfrak}
\end{align}
and used $\p_\omega\frak{D}_{o}(0,x)=\p_\omega D_i(0,x)=1/2$ (see Appendix~\ref{app:dD}). Inserting these collision integrals into the kinetic Eqs.~(\ref{KE-model2}), we end up with the coupled Fokker-Planck equations,
\begin{subequations}
\begin{align}
\p_xf_{i}(E,x)
&=
\eta \frac{N_b}{\rho_o}
\p^2_Ef_{i}(E,x)
\label{eq:12}\\
&\hspace{1mm}
+\!\eta
\Big\{[1-2f_{i}(E,x)]\p_Ef_{i}(E,x) 
+\frak{D}_{o}(0,x)\p_E^2f_{i}(E,x)\Big\},\nonumber\\
%%%%%%
\p_x \fo(E,x)&=
\label{eq:14} 
\eta
\Big\{[1-2\fo(E,x)]\p_E\fo(E,x)+D_{i}(0,x)\p_E^2\fo(E,x)\Big\},
\\
%%%%%
\p_x b(E,x)
&= \eta
\Big\{\p_Eb(E,x)+D_{i}(0,x)\p_E^2b(E,x)\Big\},
\label{eq:13}
\end{align}
\label{eq:Full-FP-model2}% DO NOT REMOVE %
\end{subequations}
in the limit of weakly broken translational symmetry. As stated above, these equations are valid when $\Delta E$ is the smallest energy scale, i.e.~smaller than the  variations in energy of  $b$, $f_i$, $\fo$, $D_{i}$ and $\frak{D}_{o}$. The dynamical diffusion parameters are $\eta D_{i}(0,x)$ [as in the previous model Eq.~(\ref{eq:30})] and  $\eta(N_b/\rho_o+\frak{D}_{o}(0,x))$ for the outer and inner ES, respectively. 

Evidently, including the outer ES's Fermi sea $\fo$ complicates the coupled Fokker-Planck Eqs.~(\ref{eq:Full-FP-model2}) compared to the ones without it in Eqs.~(\ref{eq:25}) and (\ref{eq:30}). Therefore, no immediate analytic solution of the coupled Fokker-Planck Eqs.~(\ref{eq:Full-FP-model2}) for \emph{all} three distributions are in sight. Nevertheless, the EDIE \emph{can} in fact be found analytically, since Eq.~(\ref{eq:13}) is formally identical to the Fokker-Planck Eq.~(\ref{eq:30}) in the absence of the Fermi sea of the outer ES. Consequently, the average energy $\langle E\rangle$ of the EDIE with any initial form still follows Eq.~(\ref{eq:averge-E}) and  for an initial Gaussian EDIE, the solution is still
\begin{equation}
b(E,x)=\frac{N_b}{\Gamma(x)\rho_o\sqrt{\pi}}
\exp\!\left[-\frac{[E-E_0(x)]^2}{\Gamma(x)^2}\right]
\label{eq:gauss-EDIE-model2}
\end{equation}
as in Eq.~(\ref{eq:15}), where $E_0(x)$ and $\Gamma(x)$ still follow Eqs.~(\ref{eq:max-gauss}) and (\ref{eq:4}), respectively. However, the smearing of the inner ES's Fermi sea, $D_i(0,x)$, is now \emph{different} from the one found previously in Eq.~(\ref{eq:28}), which affects the width of the EDIE only. In fact, from the point of view of the EDIE, the present extension of the previous model in Sec.~\ref{sec:coupl-fokk-planck} only serves to refine the modelling of $D_i(0,x)$. Below, we will model $D_i(0,x)$ within an effective temperature approach. 

The difficulty of solving the coupled Fokker-Planck Eqs.~(\ref{eq:Full-FP-model2}) is partly due to the \emph{non-linear} drift terms $(1-2f_i)\p_Ef_i$ and $(1-2\fo)\p_E\fo$. These stem from the fact that $f_i$ and $\fo$ are not always small, so the Pauli-blocking terms $f_i(1-f_i)$ and $(1-\fo)\fo$ \emph{cannot} be simplified in the collision integrals (\ref{eq:coll-int-Fermi-seas}). Such non-linear terms makes the Fokker-Planck scheme less fruitful for an initial step distribution as previously studied by other methods theoretically \cite{Lunde-2010} and  experimentally.\cite{Altimiras-Nat-phys-2010,le-Sueur-Altimiras-PRL-2010,Altimiras-2010a,Otsuka-et-al-2014}
  
To illustrate the differences between the models with and without a Fermi sea in the outer ES, we provide two simulation movies in the supplementary material.\cite{supp-mat} They compare our two models together with a numerical solution of the full kinetic Eqs.~(\ref{full-KE}) in the Fokker-Planck regime (named \texttt{model\_compI.avi}) and outside the Fokker-Planck regime (called \texttt{model\_compII.avi}). The latter regime will be discussed in Sec.~\ref{sec:beyond-FP}. Both simulations clearly show  an overheating of the inner ES in the model without the outer ES's Fermi sea as expected. Moreover, the Fokker-Planck models are found to compare very well to the full kinetic Eqs.~(\ref{full-KE}) within the Fokker-Planck regime. 

\subsection{Conserved quantities within this model}\label{conserved-quan-model2}

Now we show that the coupled Fokker-Planck Eqs.~(\ref{eq:Full-FP-model2}) fulfil the conservation laws in Sec.~\ref{sec:conserved-quan}, so they are a sensible approximation to the full coupled kinetic Eqs.~(\ref{full-KE}). 

First of all,  the coupled  Fokker-Planck Eqs.~(\ref{eq:Full-FP-model2}) conserve the particle number of the Fermi seas $f_i$ and $\fo$ and the EDIE $b$ separately, since
\begin{align}
\p_x\!\int_{\epsilon_c}^{\infty}\!\!\! d E f_{i}(E,x)
&=
\p_x\!\int_{\epsilon_c}^{\infty}\!\!\! d E \fo(E,x)
=
\p_x\!\int_{-\infty}^{\infty}\!\!\! d E b(E,x)
=0.
\end{align}
This is shown by inserting each of the three Fokker-Planck Eqs.~(\ref{eq:Full-FP-model2}) and using $(1-2g)\p_Eg=\p_E[(1-g)g]$ for $g=f_i,\fo$ as well as the appropriate high and low energy limits of the distributions and their derivatives.  

Secondly, we confirm the conservation law 
\begin{align}
\p_x[Z_{f_i}(x)+Z_{\fo}(x)+Z_{b}(x)]=0,
\label{eq:energy-conservation-model2}
\end{align} 
expressing energy conservation in the scattering. Due to our partitioning of the outer distribution, $f_o=\fo+b$, we have here divided $Z_o(x)$ for the outer ES into two parts as $Z_o(x)=Z_{\fo}(x)+Z_{b}(x)$. Therefore, here we label $Z$ by the distribution instead of the ES as in Eq.~(\ref{energy-conservation-general}), i.e.
\begin{align}
Z_{g}(x)&=\int_{-\infty}^{\infty} d E\; (E-\bar\mu_{g})g(E,x),
\label{eq:def-Z-g}
\end{align}
for $g=b,\fo,f_i$. From the Fokker-Planck Eqs.~(\ref{eq:Full-FP-model2}) and the high and low energy limits of the distributions, we find (see Appendix \ref{app:dZ} for details)
\begin{subequations}
\begin{align}
\p_x Z_b(x)&=-\eta \frac{N_b}{\rho_o},
\label{eq:dZb-model2}\\
\p_x Z_{f_i}(x)&=\eta \frac{N_b}{\rho_o}-\eta D_i(0,x)+\eta\frak{D}_o(0,x),
\label{eq:dZfi-model2}\\
\p_x Z_{\fo}(x)&=-\eta \frak{D}_o(0,x)+\eta D_i(0,x),
\label{eq:dZfo-model2}
\end{align}
\label{eq:Zs-model2}%HERE THE "%" RIGHT AFTER THE LABEL MUST NOT BE REMOVED!!!
\end{subequations}
so the conservation law in Eq.~(\ref{eq:energy-conservation-model2}) is fulfilled.  We observe from Eq.~(\ref{eq:dZb-model2}) that the injected electrons lose energy at a constant rate, $\propto \eta N_b$, as in the model without the outer ES's Fermi sea [compare to Eq.~(\ref{eq:dZ-model1})]. According to  Eq.~(\ref{eq:dZfi-model2}), the inner ES's Fermi sea $f_i$ absorbs this energy at the same rate, $\eta N_b/\rho_o$, just like in Eq.~(\ref{eq:dZ-model1}). However, now the Fermi seas of the inner and outer ESs also exchange energy with rates proportional to their energy smearing,  $D_i(0,x)$ and $\frak{D}_o(0,x)$, respectively, as evident from Eqs.~(\ref{eq:dZfi-model2}) and (\ref{eq:dZfo-model2}). This interpretation of Eq.~(\ref{eq:Zs-model2}) motivate us to construct an effective temperature approach below in order to better understand the physics of the Fermi sea distributions $f_i$ and $\fo$ in the relaxation. 

\subsection{Effective temperature approach\label{sec:effect-temp-appr}}

Up to now, we have several times discussed the energy smearing around the inner and outer Fermi levels, $D_i(0,x)$ and $\frak{D}_o(0,x)$, and intuitively understood these as temperatures of the two Fermi seas ---although we are in an out-of-equilibrium situation. Here we take this intuition one step further and gain transparent physical insights into the relaxation of the injected electrons in the Fokker-Planck regime described by Eqs.~(\ref{eq:Full-FP-model2}). To this end, we use the Ansatz that the Fermi seas are simply distributed according to Fermi functions with $x$-dependent effective temperatures $T_i(x)$ and $T_o(x)$, i.e. 
\begin{subequations}
\begin{align}
f_{i}(E,x)&\rightarrow
f^F_{i}(E,x)
\equiv
\frac{1}{1+\exp\!\Big[\frac{E-\mu_{i}}{\kb T_{i}(x)}\Big]},\label{eq:16} \\ 
%%%%%%%%%%
\fo(E,x)&\rightarrow
\fo^F(E,x)
\equiv
\frac{1}{1+\exp\!\Big[\frac{E-\mu_{o}}{\kb T_{o}(x)}\Big]}.\label{eq:17}
\end{align}
\label{eq:Fermi-ansatz-model2}%   %HERE THE "%" RIGHT AFTER THE LABEL MUST NOT BE REMOVED!!!
\end{subequations}
Inserting these into the definitions (\ref{eq:20}) and (\ref{eq:def-Dfrak}) promptly gives
\begin{align}
D_i^F(0,x)=\kb T_i(x)
\quad\textrm{and}\quad
\frak{D}^F_o(0,x)=\kb T_o(x), 
\end{align}
where the superscript $F$ indicates the use of the Ansatz (\ref{eq:Fermi-ansatz-model2}). Thus, the energy smearing around the inner and outer ES's Fermi level is simply the effective temperature, also out-of-equilibrium for $x>0$. This is the basic idea of our effective temperature approach. We emphasize that Eq.~(\ref{eq:Fermi-ansatz-model2}) is \emph{not} the exact solution to the coupled Fokker-Planck Eqs.~(\ref{eq:Full-FP-model2}), but it is a good approximation as discussed below. Furthermore, we stress that the effective temperatures are \emph{not} to be understood as strict thermal equilibrium quantities, but merely as the energy smearing of the Fermi seas.    

We use the energy conservation in the relaxation as expressed by Eqs.~(\ref{eq:Zs-model2}) to build the basic equations for the  effective temperatures $T_i(x)$ and $T_o(x)$. The Ansatz (\ref{eq:Fermi-ansatz-model2}) allows us to evaluate the left-hand sides of Eqs.~(\ref{eq:dZfi-model2}) and (\ref{eq:dZfo-model2}), i.e. 
\begin{align}
\p_xZ^{F}_{f_i}(x)&= 
\int^{\infty}_{-\infty} \!\!\!\!d E (E-\mu_{i}) \p_x f_i^F(E,x)
=
\frac{\pi^2}{3^{}} 
\kb T_{i}(x) \kb \p_xT_{i}(x) ,
\end{align} 
and $\p_xZ^{F}_{\fo}(x)=\frac{\pi^2}{3}\kb T_{o}(x) \kb \p_xT_{o}(x)$, where we chose $\bar\mu_{f_i}=\mu_i$ and $\bar\mu_{\fo}=\mu_o$. Inserting these into Eqs.~(\ref{eq:dZfi-model2}) and (\ref{eq:dZfo-model2}) leads to the basic equations for the evolution of $T_i(x)$ and $T_o(x)$ as
\begin{subequations}
\label{eq:eff-temp-diff-eq}
\begin{align}
\kb \p_xT_{i}(x) 
&= 
\eta
\frac{3}{\pi^2}
\left(
\frac{N_b/\rho_o}{\kb T_{i}(x)}+\frac{T_{o}(x)}{T_{i}(x)}-1
\right),
\label{eq:19}\\
\kb \p_xT_{o}(x)&= 
\eta
\frac{3}{\pi^2}
\left(
\frac{T_{i}(x)}{T_{o}(x)}-1
\right).
\label{eq:18}
\end{align}
\end{subequations}
These equations describe the relaxation of the Fermi seas as long as the EDIE and the Fermi sea are well separated in energy, just as for the Fokker-Planck Eqs.~(\ref{eq:Full-FP-model2}). The term $N_b/(\rho_o\kb T_{i})$ originates from the collision integral between the EDIE and $f_i$, while the terms $T_{i}/T_{o}-1$ and $T_{o}/T_{i}-1$ stem from the collision integrals connecting the two Fermi seas. 

Now we explain how the effective temperature Eqs.~(\ref{eq:eff-temp-diff-eq}) give a physically transparent picture of the relaxation as illustrated in Fig.~\ref{fig:1}.  At the injection point $x=0$, $T_i(0)=T_o(0)=T$ and according to Eq.~(\ref{eq:18}), $\p_x T_o(0)=0$, while $\kb \p_xT_i(0)=\eta\frac{3}{\pi^2}\frac{N_b}{\rho_o\kb T}>0$ due to Eq.~(\ref{eq:19}). Thus initially, the injected electrons relax by heating up the inner ES \emph{only} as seen on Fig.~\ref{fig:1}(a-b). Therefore, $T_i(dx)>T_o(dx)$ for some small distance $dx$.  This, in turn, forces $T_o(x)$ to increase since $T_{i}(dx)/T_{o}(dx)-1>0$ in Eq.~(\ref{eq:18}), while the rate of heating of the inner ES [i.e. the right-hand side of  Eq.~(\ref{eq:19})] is reduced. This is exactly the energy redistribution between the Fermi seas shown in Fig.~\ref{fig:1}(c-d), where the Fermi sea in the outer ES absorbs some of the heat from the inner ES. It is now evident that $T_i(x)>T_o(x)$ throughout the relaxation and that it is the relaxation of the injected electrons that drives the heating through the term $N_b/(\rho_o\kb T_{i}(x))$ in Eq.~(\ref{eq:19}).\cite{footnote-diff-ini-temps}

\begin{figure}
\includegraphics[width=\columnwidth]{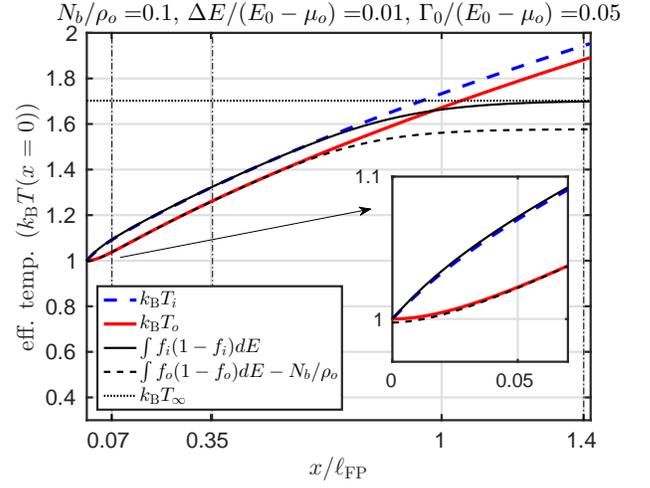}
\caption{(Color online) The effective temperatures $T_o(x)$ (full red curve) and $T_i(x)$ (dashed blue curve) according to  Eqs.~(\ref{eq:eff-temp-diff-eq}) compared with $D_i(0,x)$ and $D_o(0,x)-N_b/\rho_o\simeq \mathfrak{D}_o(0,x)$ (full and dashed black curves) obtained by numerical iteration of the full kinetic equations (\ref{full-KE}). These quantities are shown in units of the initial device temperature $\kb T$. The dotted horizontal line gives the temperature $T_{\infty}$ of the fully relaxed distributions Eq.~(\ref{eq:T-infty}). The dotted-dashed vertical lines give the positions of the distribution snapshots in Fig.~\ref{fig:distr_iter}. The inset shows a zoom-in of the short distance behavior. The parameters are in the Fokker-Planck regime, i.e.~$\Delta E$ is the smallest energy scale and the EDIE fulfills $b(E,x)\ll 1$. The effective temperature Eqs.~(\ref{eq:eff-temp-diff-eq}) only depend on $\eta=\sqrt{\pi}\gamma (\Delta E)^3/4=0.014$ and $N_b/\rho_o=0.1$, whereas the full kinetic Eqs.~(\ref{full-KE}) use $\Delta E=0.4$, $\gamma=0.5$, $\kb T=0.8$, $E_0-\mu_o=40$ and $\Gamma_0=2$ (in arbitrary units) and an initial Gaussian EDIE.  
\label{fig:iter_eff_temp2}}
\end{figure}

Fig.~\ref{fig:iter_eff_temp2} presents the effective temperatures found numerically from the system of differential equations (\ref{eq:eff-temp-diff-eq}). This numerical example agrees with the general evolution of the effective temperatures discussed above. In particular, the inner ES heats up rapidly for small $x$ as seen in the inset, whereas the outer ES needs some temperature difference, $T_i(x)-T_o(x)$, before it begins to heat up as well. In Fig.~\ref{fig:iter_eff_temp2}, the effective temperatures are compared to the energy smearing of the Fermi  levels, $D_i(0,x)$ and $D_o(0,x)-N_b/\rho_o$, as obtained by an iterative numerical solution of the full kinetic  equations~(\ref{full-KE}). The  comparison is nearly perfect in the regime of validity of our Fokker-Planck approach in                                                                                      Sec.~\ref{subsec:FP-model2}, i.e. when  $\Delta E$ is the smallest energy scale, $b(E,x)\ll 1$,  and the EDIE is well separated  in energy from the Fermi sea $\fo$, which is valid for $x<\ell_{\textsc{fp}}$.  In this regime, we have $D_o(0,x)-N_b/\rho_o\simeq \frak{D}_o(0,x)$. Here $D_o(0,x)$ is calculated using the \emph{entire} outer ES distribution from the full kinetic equations, whereas $\frak{D}_o(0,x)$ is for the Fermi sea distribution only. Note that the full kinetic Eqs.~(\ref{full-KE}) do not give direct access to the outer ES's Fermi sea distribution separated from the EDIE, see Appendix \ref{app:compare-eff-temp-and-model2}. Physically, the heating of the ESs stops once  the Fermi sea absorbs the injected electrons for $x\gtrsim \ell_{\textsc{fp}}$ and the system thereby fully relaxes. This is not captured by our effective temperature model (evident from Fig.~\ref{fig:iter_eff_temp2} and Eq.~(\ref{eq:cons-law-eff-temp}) below), since it models the case of the EDIE $b$ and Fermi sea  $\fo$ being separated in energy valid for $x<\ell_{\textsc{fp}}$. However, the entire relaxation is captured by the full kinetic Eqs.~(\ref{full-KE}) as seen on Fig.~\ref{fig:iter_eff_temp2}, where both $D_i(0,x)$ and $D_o(0,x)$ go to the asymptotic temperature $\kb T_{\infty}$ of the fully relaxed system to be discussed in the next section [see Eq.~(\ref{eq:T-infty})].    

\begin{figure*}
\includegraphics[width=0.67\columnwidth]{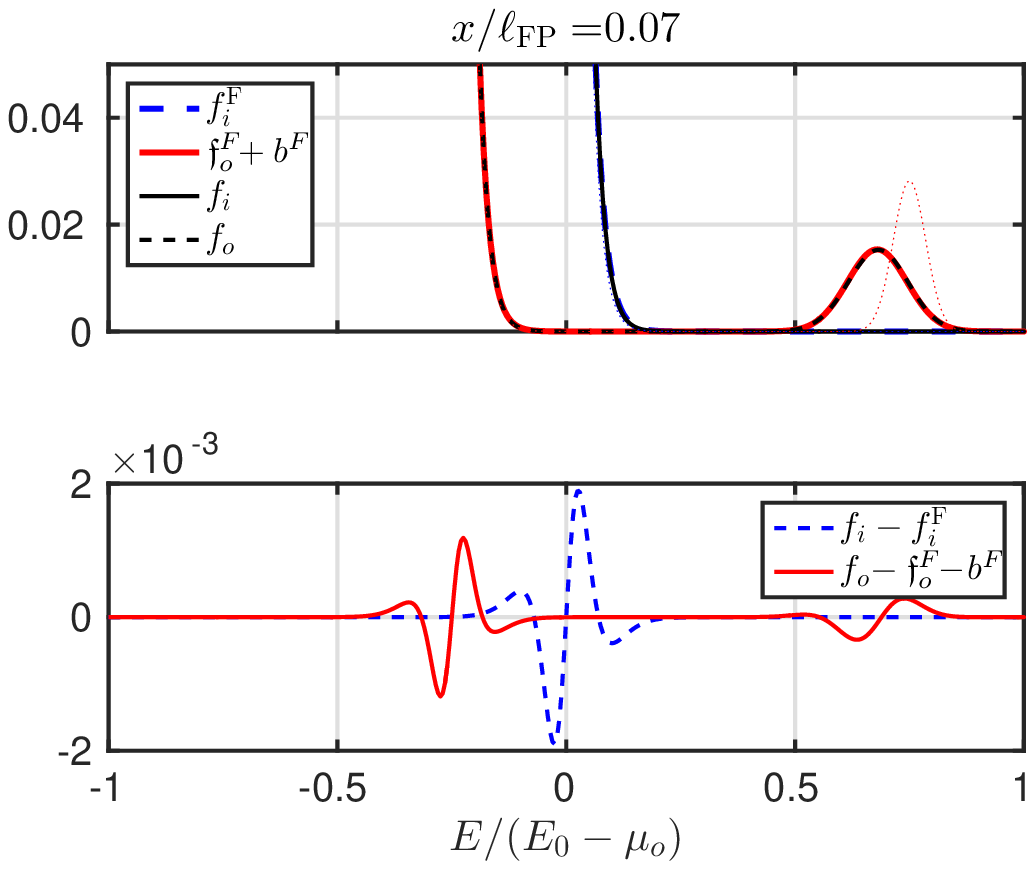}\hfill
\includegraphics[width=0.67\columnwidth]{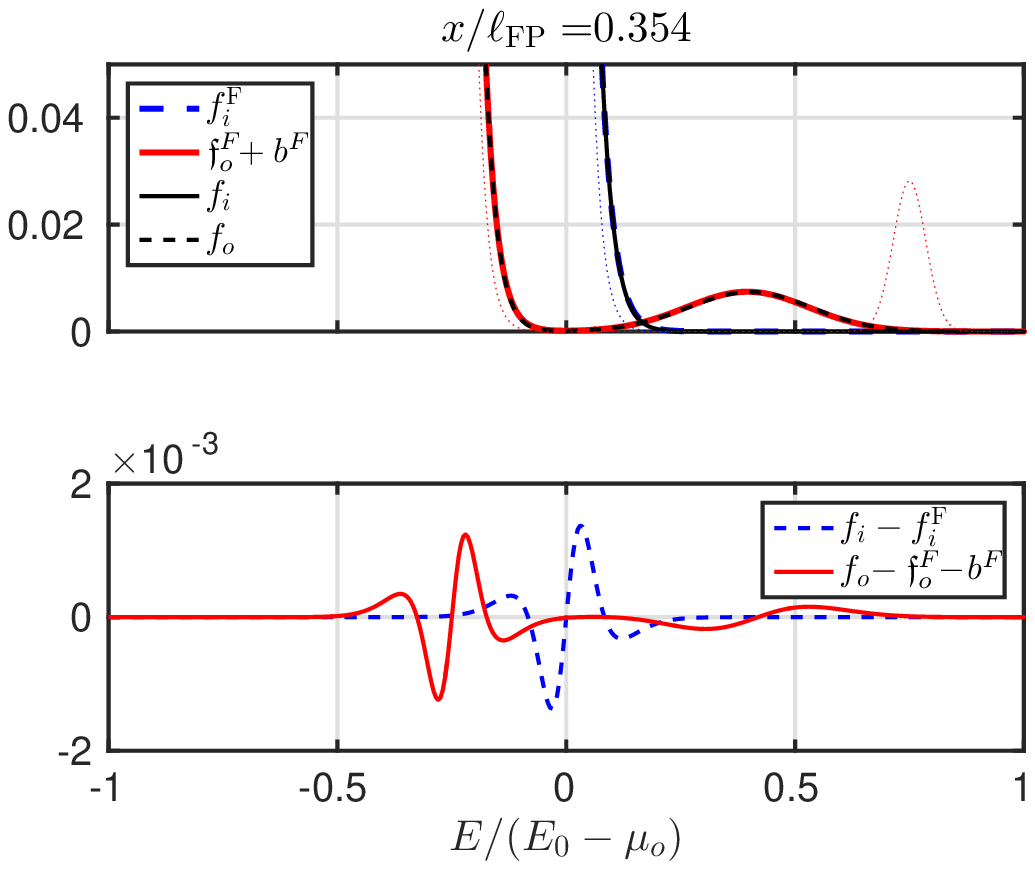}\hfill 
\includegraphics[width=0.67\columnwidth]{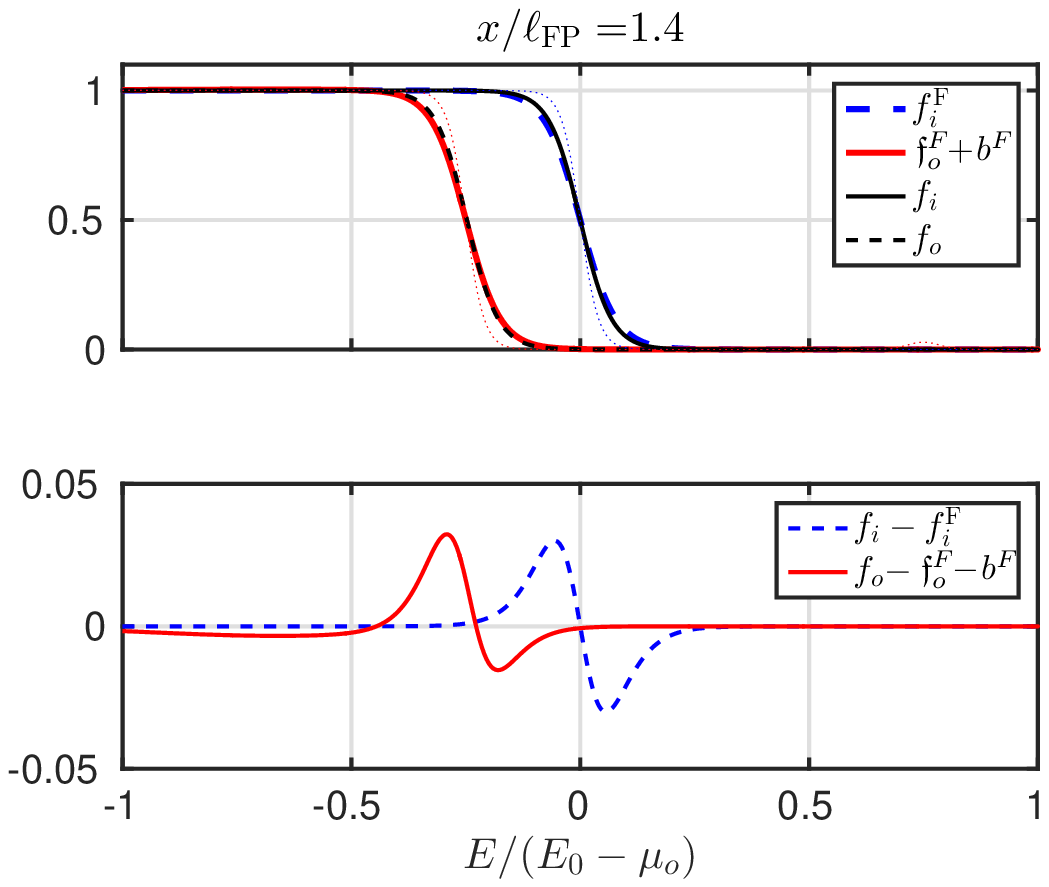}
\caption{(Color online) The distribution functions of the inner and outer ESs at distances $x=0.07\ell_{\textsc{fp}}$, $x=0.354\ell_{\textsc{fp}}$ and $x=1.4\ell_{\textsc{fp}}$ (indicated by vertical dashed-dotted lines in Fig.~\ref{fig:iter_eff_temp2}). These are obtained by (i) numerical iteration of the full kinetic Eqs.~(\ref{full-KE}) ($f_i$ and $f_o$ shown as full and dashed black curves) and (ii) from the effective temperature Fermi Ansatz distributions $f_i^F$ and $\fo^F$ Eq.~(\ref{eq:Fermi-ansatz-model2}) combined with the Gaussian EDIE $b^F$ Eq.~(\ref{eq:gauss-EDIE-model2}) with $D_i(0,x)$ exchanged by $\kb T_i(x)$ in $\Gamma(x)$ in Eq.~(\ref{eq:4}) (shown by dashed blue and full red curves).  The thin dotted (blue and red) curves indicate the initial distributions. Note the scale change of the vertical axis between the figures for $x<\ell_{\textsc{fp}}$ and $x>\ell_{\textsc{fp}}$. The difference between the distributions found by the two approaches are given in the lower panels. The difference increases significantly for $x\gtrsim \ell_{\textsc{fp}}$, since it is outside the regime of validity of our Fokker-Planck approach. All parameters are identical to those of Fig.~\ref{fig:iter_eff_temp2}. 
\label{fig:distr_iter}}
\end{figure*}

Fig.~\ref{fig:distr_iter} shows three snapshots of the distribution functions for the parameters of Fig.~\ref{fig:iter_eff_temp2} at distances $x=0.07\ell_{\textsc{fp}}$, $x=0.354\ell_{\textsc{fp}}$ and $x=1.4\ell_{\textsc{fp}}$ (vertical dashed-dotted lines in Fig.~\ref{fig:iter_eff_temp2}). The effective temperature Fermi Ansatz distributions (\ref{eq:Fermi-ansatz-model2}) are found to compare well with the results of the full kinetic Eqs.~(\ref{full-KE}) for  $x<\ell_{\textsc{fp}}$, which is evident from the difference between the two approaches (shown below the distributions in Fig.~\ref{fig:distr_iter}). Outside the regime of validity of the Fokker-Planck approach, $x\gtrsim\ell_{\textsc{fp}}$, a larger difference is found as expected.

The effective temperature approach was constructed to approximate the coupled Fokker-Planck Eqs.~(\ref{eq:Full-FP-model2}) and therefore, we present a separate comparison between them in Appendix \ref{app:compare-eff-temp-and-model2} and Fig.~\ref{fig:compare-eff-temp-model2}. This shows that the effective temperature approach contains the correct physics for $x<\ell_{\textsc{fp}}$, even for larger values of $N_b/\rho_o$ than presented in Fig.~\ref{fig:iter_eff_temp2}.

Now we present a few analytical insights into the effective temperatures. Firstly, by calculating $\p_x[T_i(x)^2]+\p_x[T_o(x)^2]$ from Eqs.~(\ref{eq:eff-temp-diff-eq}) and integrating over $x$, we get the exact relation   
\begin{align} 
\big[\kb T_i(x)\big]^2+\big[\kb T_o(x)\big]^2=2(\kb T)^2+\frac{6}{\pi^2} \frac{N_b}{\rho_o}\eta x.
\label{eq:cons-law-eff-temp}
\end{align}
I.e.~the sum of the squared effective temperatures  increase linearly with $x$. Secondly, we obtain an approximate solution for $T_i(x)$ and $T_o(x)$ in Appendix \ref{app:asym-eff-temp}, which is based on the observation that $[T_i(x)-T_o(x)]/[T_i(x)+T_o(x)]$ is small and decreases for increasing $x$. For intermediate values of $x$, well beyond the initial heating of mainly the inner ES and before full relaxation, $x<\ell_{\textsc{fp}}$, this approximation shows that 
\begin{subequations}
\label{eq:approx-eff-temp-text}
\begin{align} 
\kb T_i(x)
&\simeq
\sqrt{(\kb T)^2+\frac{3N_b \eta  x}{\pi^2\rho_o}}
+\frac{N_b}{4\rho_o},
\\
\kb T_o(x)
&\simeq
\sqrt{(\kb T)^2+\frac{3N_b \eta  x}{\pi^2\rho_o}}
-\frac{N_b}{4\rho_o}.
\end{align}  
\end{subequations}
These satisfy Eq.~(\ref{eq:cons-law-eff-temp}) to lowest order in $N_b/\rho_o$, i.e.~in the Fokker-Planck regime with a small EDIE. Thus, the difference of the effective temperatures is found to be simply $N_b/(2\rho_o)$  in the intermediate regime as observed in Fig.~\ref{fig:iter_eff_temp2}. To reach this regime, the injection energy $E_0$ has to be sufficiently large.     

Finally, we point out that the effective temperature approach for the model \emph{without} the outer ES's Fermi sea in Sec.~\ref{sec:coupl-fokk-planck}, gives $\kb \p_xT_{i}(x)=\eta\frac{3}{\pi^2}\frac{N_b}{\rho_o\kb T_{i}(x)}$  such that $\kb T_{i}(x)=\sqrt{(\kb T)^2+6N_b\eta x/(\pi^2\rho_o)}$. Comparing with Eq.~(\ref{eq:cons-law-eff-temp}), it is clear that the model without the outer ES's Fermi sea produces an overheated inner ES. Including both Fermi seas, the heating is redistributed between the ESs. 

\section{The fully relaxed state}\label{sec:full-relax}

At first the injected electrons gradually lose energy ($x<\ell_{\textsc{fp}}$), then they get absorbed by the Fermi sea ($x\sim\ell_{\textsc{fp}}$) and finally the ESs reach a fully relaxed state ($x\gg\ell_{\textsc{fp}}$). Now we discuss the final fully relaxed state. Numerical iteration of the full kinetic Eqs.~(\ref{full-KE}) leads to a fully relaxed state consistent with Fermi distributed ESs, i.e.
\begin{subequations}
\begin{align}
f_o(E,x\gg\ell_{\textsc{fp}})&=\frac{1}{1+\exp\left[{\frac{E-\mu_o^{\infty}}{\kb
      T_{\infty}}}\right]},\\
f_i(E,x\gg\ell_{\textsc{fp}})&=\frac{1}{1+\exp\left[\frac{E-\mu_i^{\infty}}{\kb T_{\infty}}\right]}.
\end{align}
\label{full-relax-Fermi-functions}%   %HERE THE "%" RIGHT AFTER THE LABEL MUST NOT BE REMOVED!!! 
\end{subequations}
These distributions indeed solve the full kinetic Eqs.~(\ref{full-KE}), since the collision integral (\ref{eq:2}) of equal temperature Fermi functions is zero. We determine the Fermi levels, $\mu_o^{\infty}$  and $\mu_i^{\infty}$, and the temperature $T_{\infty}$ of the fully relaxed distributions by using the conservation laws discussed in Sec.~\ref{sec:conserved-quan}. 
First of all, the particle number is conserved in each ES separately, i.e. 
\begin{equation}
\label{eq:particle-cons}
\int^\infty_{\epsilon_c} dE f_\alpha(E,x=0)
=\int^\infty_{\epsilon_c} dE f_\alpha(E,x\gg\ell_{\textsc{fp}})
\end{equation}
for $\alpha=i,o$. Inserting Eq.~(\ref{full-relax-Fermi-functions}) and the initial distributions give
\begin{equation}
\mu_i^{\infty}=\mu_i 
\quad \textrm{and}\quad
\mu_o^{\infty}=\mu_o+\frac{N_b}{\rho_o},
\end{equation}
i.e.~absorbing the injected electrons naturally increase the Fermi level of the outer ES. Secondly, according to Eq.~(\ref{energy-conservation-general}) the energy conservation in the scattering implies that 
\begin{equation}
Z_i(x=0)+Z_o(x=0)=Z_i(x\gg\ell_{\textsc{fp}})+Z_o(x\gg\ell_{\textsc{fp}}), 
\end{equation}
where we choose $\bar\mu_\alpha=\mu_\alpha$ for convenience. This leads to
\begin{equation}
\label{eq:T-infty}
\kb T_{\infty}=
\sqrt{(\kb T)^2
+\frac{3}{\pi^2}
\frac{N_b}{\rho_o}
\left[\Big(E_0-\mu_o\Big)-\frac{1}{2}\frac{N_b}{\rho_o}\right]},
\end{equation}
which is obtained {\em solely} from the initial and the fully relaxed  distributions,  i.e.~$T_{\infty}$, $\mu_i^{\infty}$ and $\mu_o^{\infty}$ are  independent of the assumptions used to derive the Fokker-Planck dynamics. Here $E_0$ is the initial average energy of the injected electrons, i.e.~$E_0=\langle E\rangle_{x=0}$ [see Eq.~(\ref{eq:def-average-E})], which is equal to the initial maximum for a symmetric EDIE (see Appendix \ref{app:symmetric-packet}). Fig.~\ref{fig:iter_eff_temp2}  shows how $\kb T_{\infty}$ is approached by $\int dE f_\alpha (1-f_\alpha)$ for $\alpha=i,o$ found from the full kinetic equations as $x$ increases beyond $\ell_{\textsc{fp}}$. Moreover, when calculating the particle number (\ref{eq:particle-cons}) as well as $Z_\alpha$ for the Fermi distributions, it is necessary to choose the cut-off energy $\epsilon_c$ sufficiently low to obtain the above  expressions. As expected, the final results are independent of $\epsilon_c$.

In the previous sections, we have studied electrons injected high above the Fermi level such that the Fermi seas of the ESs are heated as a result of the relaxation, i.e.~$T_{\infty}>T$. Remarkably, the opposite $T_{\infty}<T$ is actually also possible, namely if the electrons are injected close to the Fermi level such that $E_0-\mu_o <N_b/(2\rho_o)$ as seen from Eq.~(\ref{eq:T-infty}). In this case, the \emph{inner ES is therefore cooled down} compared to the initial temperature. This might seem counterintuitive at first, since the initial average energy of the injected electrons is higher than the Fermi level, $E_0-\mu_o>0$. However, in this case the Fermi sea and the EDIE in the outer ES have to a large extend merged such that the temperature of the outer ES's Fermi sea  looses its physical meaning.\cite{footnote-T_inf-lower-than-T} Instead, the initial out-of-equilibrium distribution of the outer ES pictorially appears colder, i.e.~with a steeper transition from one to zero than $\kb T$. This is therefore physically comparable to the case of two ESs with different initial temperatures, where the hotter ES also will cool down in the relaxation process. Therefore, it is now clear that $T_{\infty}<T$ is indeed physically possible. Interestingly, it could be used proactively to cool down an ES below the device temperature by electrostatic means in future experiments.   

\section{Discussion}\label{sec:discussion}

\subsection{Relaxation beyond the Fokker-Planck regime}\label{sec:beyond-FP} In the preceding we have shown that under the conditions that (i) the occupancies of the high energy states remain low, $b(E,x)\ll1$, and (ii) that $\Delta E$ characterizing the energy exchange per collision is the smallest energy scale, then the relaxation can be described as a generalized drift-diffusion process in energy modelled  by a Fokker-Planck equation. Physically this corresponds to an injection QD weakly coupled to an ES, along which the translational invariance is only weakly broken. In this section, using the numerical solution of the full kinetic equations~(\ref{full-KE}), we explore how relaxation proceeds when either of these two conditions are not satisfied, that is relaxation beyond the Fokker-Planck regime.

\begin{figure}[ht]
\includegraphics[width=0.85\columnwidth]{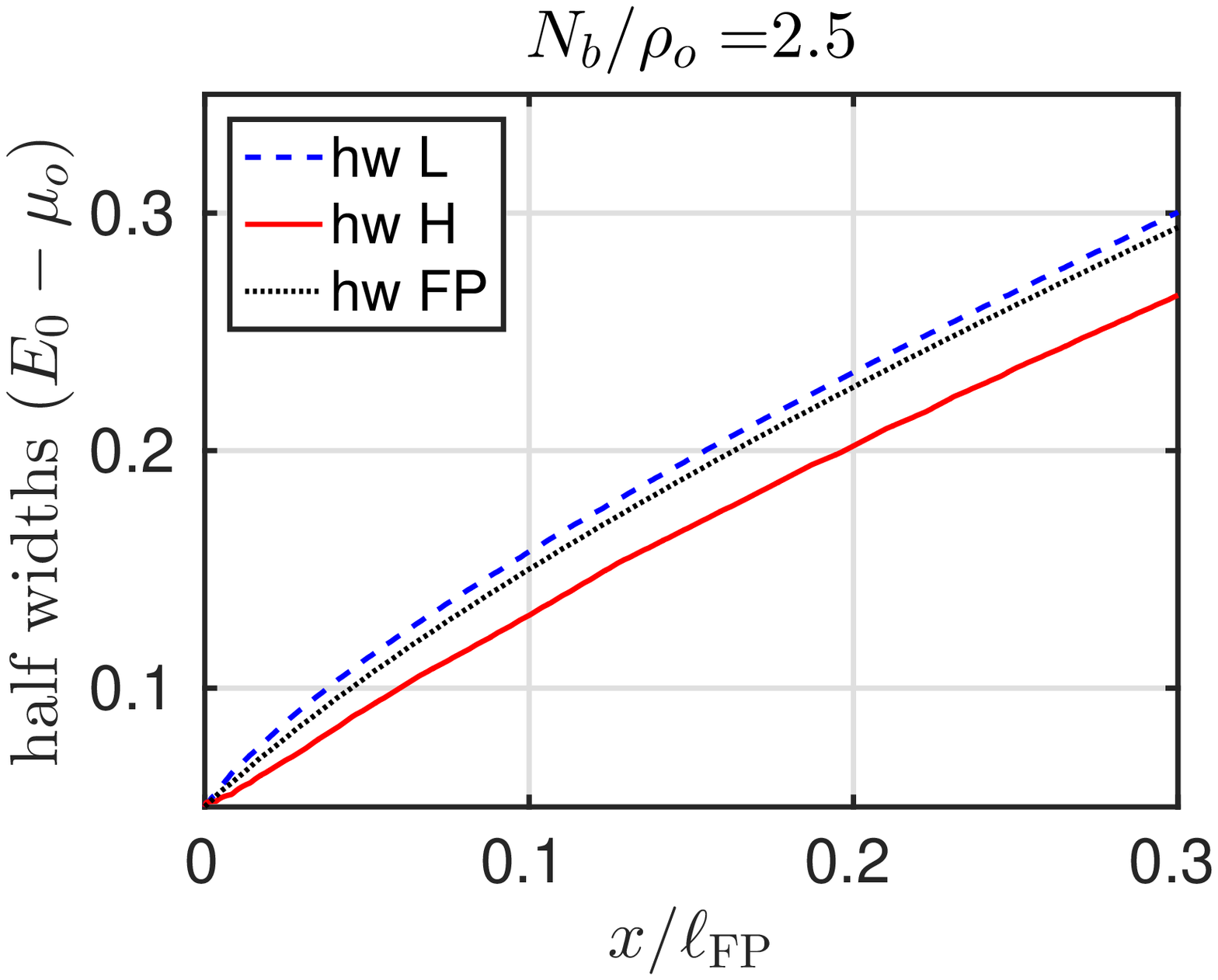}
\includegraphics[width=0.85\columnwidth]{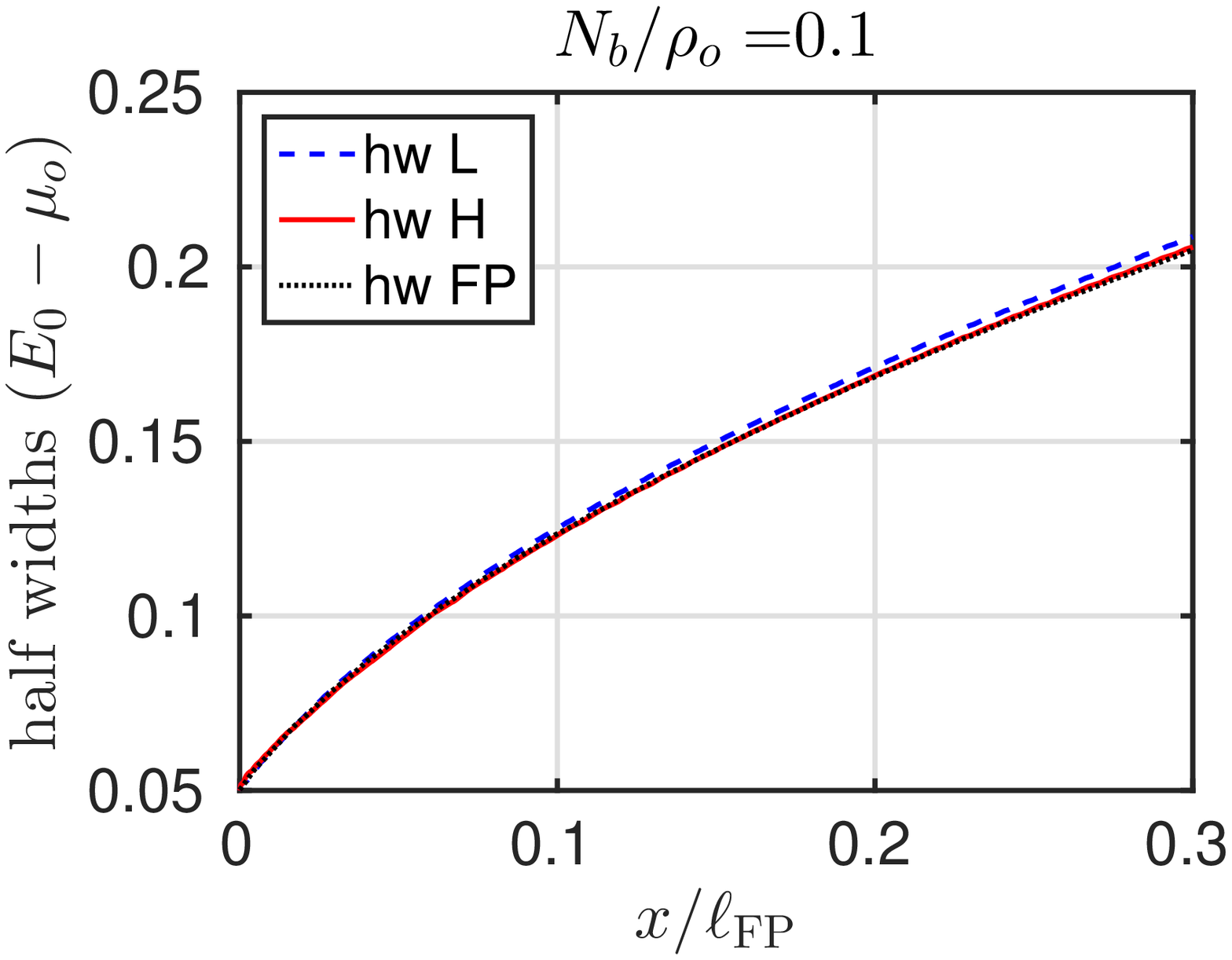}
\caption{(Color online) Symmetric vs.~asymmetric relaxation for $N_b/\rho_o=0.1$ (bottum panel) and $N_b/\rho_o=2.5$ (top panel). The low-energy half-width (hw L, dashed blue lines) and high-energy half-width (hw H, solid red lines) are obtained from the numerical solution of the full kinetic equations~(\ref{full-KE}). For comparison, we also show the half-width obtained from the coupled Fokker-Planck Eqs.~(\ref{eq:Full-FP-model2}) named hw FP  (dotted black lines). The parameters used are $\kb T/(E_0-\mu_o)=0.03$, $\Delta E/(E_0-\mu_o)=0.01$ and $\Gamma_0/(E_0-\mu_o)=0.05$.
\label{fig:asym_relax}}
\end{figure}

If the injection QD is not weakly coupled to the ES, then condition (i) is not satisfied, i.e. the occupancies of the high energy states may approach unity. However, if $\Delta E$ remains the smallest energy scale, then we may expect that modifications mainly arise due to the Pauli principle, since Pauli-blocking can no longer be neglected in the collision integrals, i.e.~$b(1-b)\nsimeq b$ for $b\nll 1$. Therefore, non-linear terms in $b$ arise and the Fokker-Planck Eq.~(\ref{eq:13}) is replaced by 
\begin{align}
\p_x b=\eta
\Big\{\big[1-2b\big]\p_Eb+D_{i}(0,x)\p_E^2b\Big\}.
\label{eq:non-linear-FP}
\end{align}
Hence, the drift term becomes non-linear. A simple consequence is a reduced drift velocity of the EDIE, since $1-2b<1$. This effect is seen in the movie \texttt{model\_compII.avi},\cite{supp-mat} where the Fokker-Planck approaches and the full kinetic Eqs.~(\ref{full-KE}) are compared for a large EDIE ($N_b/\rho_o=2.5$). Furthermore, we can understand the effect of the Pauli-blocking on the width of the EDIE as follows.  For an initial EDIE with a well-defined single maximum $E_{\textrm{max}}(0)$, we have that  $b(E-\Delta E,x)>b(E,x)$ for $E>E_{\textrm{max}}(x)$, so Pauli-blocking will suppress transitions of electrons with energies above the maximum $E_{\textrm{max}}(x)$ to states with lower energy (of order $E-\Delta  E$) but higher occupancy. In contrast, since $b(E-\Delta E,x)<b(E,x)$ for $E<E_{\textrm{max}}(x)$, electrons with an energy below the maximum $E_{\textrm{max}}(x)$ can more frequently transition to lower energy states because the latter have lower occupancy. Thus the relaxation produces asymmetry around the maximal energy. Even if the initial EDIE is symmetric with respect to the average injection energy $E_0$, it will not remain so: Its low energy side will spread faster than its high energy side. This is indeed what we observe on the top panel of Fig.~\ref{fig:asym_relax}, where we show the low and high-energy half-widths of the EDIE for $N_b/\rho_o=2.5$. We clearly see that the low-energy half-width (hw L) grows faster than the high-energy half-width (hw H). For comparison we show the corresponding behavior in the Fokker-Planck regime for $N_b/\rho_o=0.1$ in the bottom panel of Fig.~\ref{fig:asym_relax}. The  movies \texttt{symmetric.avi} and \texttt{asymmetric.avi} in the supplementary materials\cite{supp-mat} show the corresponding evolution of the distribution functions.

If condition (i) is satisfied but not condition (ii), e.g. if the energy  exchange per collision $\Delta E$ is on the order of the injection energy $E_0$, then relaxation can proceed in a jump-like fashion and not in a drift-diffusive way. This is illustrated in movie \texttt{jumpI.avi} in the supplementary materials.\cite{supp-mat}   Note that relaxation proceeds in a similar jump-like fashion if neither condition (i) nor condition (ii) are satisfied as shown in movie \texttt{jumpII.avi} of the supplementary materials.\cite{supp-mat}   Indeed if $\Delta E$ is larger than the initial width of the EDIE, then the Pauli blocking effect discussed above is irrelevant.

\subsection{Electron-hole symmetric relaxation}

Electrons and holes injected well above/below the Fermi level into a one dimensional wire have recently been shown experimentally to relax very differently in a cleaved-edge overgrowth device.\cite{Barak-Steinberg-Glazman-Yacoby-Nat-phys-2010} This asymmetric relaxation between electrons and holes can be quantitatively understood from three-particle collisions\cite{Lunde-PRB-2007} in a non-linear band using a Boltzmann kinetic equation approach.\cite{Karzig-Glazman-von-Oppen-PRL-2010} 

Using our quantum Hall setup of two QDs connected by an ES shown in Fig.~\ref{fig:1},  we could equally well analyze the relaxation of electrons removed from the Fermi sea at an energy far below the Fermi level. In other words, we could study how the energy distribution of injected holes (EDIH) change and thereby examine the possibility of electron-hole asymmetric relaxation. \emph{However}, in the theory presented here, the injected electrons and holes relax in the same way, i.e.~we find electron-hole symmetric relaxation. Pictorially, this means that the relaxation pattern of an EDIH can be found by simply making a mirror image of an equivalent EDIE in the Fermi surface. Mathematically, the electron-hole mirror image of some distribution function $f(E,x)$ in the Fermi level $\mu$ is given by $1-f(\mu-(E-\mu),x)$. Thus, the statement of electron-hole symmetric relaxation precisely means that the outer ES's distribution $f_{o,H}(E,x)$ containing injected holes is related to the distribution $f_{o,E}(E,x)$ containing injected electrons by  
\begin{align}
1-f_{o,E}(\mu_o-(E-\mu_o),x)=f_{o,H}(E,x), 
\end{align}
where the initial electron ($E$) and hole ($H$) distributions are mirror images, i.e.~$f_{o,E}(E,0)=f_0(E)+b(E,0)$  and $f_{o,H}(E,0)=f_0(E)-b(\mu_o-(E-\mu_o),0)$ with $f_0(E)$ being the Fermi function. Some amount of electron-hole asymmetry could be introduced into our theory by using e.g.~non-linear bands. Experimentally investigating the electron-hole symmetry of relaxation could help to pinpoint the relaxation mechanism and the possible need for band curvature in modelling of ESs in the integer quantum Hall regime.

\section{Summary and outlook}
We have discussed a Fermi-liquid model for relaxation and energy exchange among quantum Hall ESs, which takes into account non-momentum conserving electron-electron interactions due to e.g.~disorder and edge roughness. Specifically, we analyze the relaxation of high energy electrons injected into one of two co-propagating ESs by a tunnel-coupled QD. We  focus on the statistical properties and study the energy distribution functions of the two ESs, since these can be mapped out experimentally by a second downstream QD (see Fig.~\ref{fig:1}). Varying the distance between the two QDs enables us to study various stages of the relaxation process. Our analysis is relevant for a steady-state situation in contrast to recent studies of a QD laterally coupled to an ES with a time-dependent component.\cite{Feve-2007,Levitov-JoMP-1996,Ivanov-PRB-1997,Keeling-PRL-2006,Nigg-PRL-2006,Battista-Samuelsson-PRB-2011,Albert-PRL-2012,Parmentier-PRB-2012,Dubois-Nature-2013,Bocquillon-Annalen-der-Physik-2014,Dasenbrook-PRL-2014,Dasenbrook-PRB-2015} By including a finite width of the EDIE, our setup is close to an actual experimental realization, where tunnel-coupling always introduces a broadening of the injected energy. The system considered here is also very closely related to the system recently realized experimentally by Tewari {\emph et al.},\cite{Tewari-PRB-2016} to investigate quantum coherence of  chiral excitations  in the integer quantum Hall regime. Our results could help understand   some   of the   more puzzling observations made in this experiment,  which clearly disagree with a number of  other theoretical descriptions.~\cite{Slobodeniuk-PRB-2016} A discussion of dephasing within  this framework is however beyond the scope of the present paper.

Here we start from the Boltzmann kinetic equation for the distribution functions of the ESs with a two-body collision integral and derive a set of coupled Fokker-Planck equations. These describe the relaxation of the injected electrons statistically as a generalized drift-diffusion process in energy space. We find a constant drift velocity of the EDIE, i.e.~the average energy of the injected electrons  decreases towards  the Fermi level with  a constant rate depending only on the effective  inter-ES interaction strength and  the energy  scale  for inter-ES energy  exchange  per collision,  $\Delta E$. Moreover,  the EDIE  widens as the  electrons  relax, while the energy smearing around the Fermi levels of the  two Fermi seas increases. The Fokker-Planck dynamics is reached, when the scale of energy exchange per collision is the smallest energy scale and the EDIE is small compared to full occupation. 

The physical picture that emerges from this analysis is as follows. The injected high energy electrons in the outer ES loose energy and initially only cause electrons in the inner ES's Fermi sea to gain energy, i.e.~to heat up. Subsequently, the heat of the inner ES is redistributed between the two Fermi seas and an intermediate regime with a constant effective temperature difference between the ESs appears. Remarkably, the effective temperature difference is only determined by the number of electrons in the EDIE. The heating and the relaxation comes to an end, when the injected electrons are finally absorbed by the Fermi sea of the outer ES. 

We provide analytical solutions of both ES's energy distribution  function in a  non-perturbative out-of-equilibrium  regime     within   a  model, where the outer   ES's Fermi sea is neglected. This gives a detailed and well-controlled picture of  the relaxation. However,  due to the absence of a Fermi  sea,  the  inner ES  is overheated within this model. To describe the  energy  redistribution between  the Fermi seas, we provide a  refined model including both ESs, in which the EDIE can still be found analytically, but the exact distributions of the Fermi seas cannot. However, we introduce an intuitive effective temperature approach to describe the energy exchange in great detail, which compares well with the full Boltzmann kinetic equations.

\section{Acknowledgements} 

We thank Alex Levchenko for a useful correspondence and especially we acknowledge useful discussions with Markus B\"uttiker in the early stage of this work. AML gratefully acknowledges financial support from the Carlsberg Foundation. SEN acknowledges financial supported by the Swiss National Science Foundation Ambizione Grant no. PZ00P$2{\_}148092$. 

\appendix

\section{Conserved quantities within the kinetic equation approach}\label{appendix:conserved}

For completeness, we include a derivation of the conserved quantities in the framework of the kinetic Eq.~(\ref{Boltzmann-eq}) for general dispersion relations. In Sec.~\ref{sec:conserved-quan}, we briefly discussed these for our interaction model and linear dispersion relations.

The current $\mathfrak{I}_\alpha=(1/L)\sum_{k}v_{k\alpha}f_\alpha(k,x)$ in a single band $\alpha$ is conserved. Here, $L$ is the length of the entire system. The conservation of current, $\p_x\mathfrak{I}_\alpha=0$, can be shown by substituting $\p_xf_{\alpha}(k_1,x)$ by the collision integral [Eq.~(\ref{Boltzmann-eq})], i.e.        
\begin{align}
\p_x\mathfrak{I}_\alpha
=&
\frac{1}{L}
\sum_{k_1^{}}v_{k_1\alpha}\p_xf_{\alpha1}
=
\frac{1}{L}
\sum_{k_1^{}}I_{k_1^{}x\alpha}[f_{\alpha},f_{\bar\alpha}]
\\
=&
\frac{1}{L}
\sum_{k_1^{}k_2^{}k_{1'}k_{2'}}
W_{1^{}2^{},{1'}{2'}}
\Big\{
f_{\alpha1'}[1-f_{\alpha1}]f_{\bar{\alpha}2'}[1-f_{\bar{\alpha}2}]
\nonumber\\
&
\hspace{2.6cm}
-\underbrace{f_{\alpha1}[1-f_{\alpha1'}] f_{\bar{\alpha}2}[1-f_{\bar{\alpha}2'}]}_{\textrm{Interchange}\ k_1\leftrightarrow k_{1'}\ \textrm{and}\ k_{2}\leftrightarrow k_{2'}}
\Big\}=0
\nonumber
\end{align}
using the short-hand notation $f_{\alpha i}=f_{\alpha}(k_i,x)$ and assuming a symmetric scattering rate: $W_{1^{}2^{},{1'}{2'}}=W_{{1'}{2'},1^{}2^{}}$. \emph{We emphasize that in the case of linear bands used in the main part of this paper, the conservation of current and number of particles in} Eq.~(\ref{particle-conservation-general}) \emph{are identical}. In the main text, we use the term conserved number of particles.  

Next, we show the conservation of the total energy current $\mathfrak{I}_E=(1/L)\sum_{k\alpha}E_{k\alpha}v_{k\alpha}f_\alpha(k,x)$ in all bands, i.e   
\begin{widetext}
\begin{align}
\p_x\mathfrak{I}_E
=&
\frac{1}{L}
\sum_{k_1^{}\alpha}E_{k_1\alpha}v_{k_1\alpha}\p_xf_{\alpha1}
=
\frac{1}{L}
\sum_{k_1^{}\alpha}E_{k_1\alpha}I_{k_1^{}x\alpha}[f_{\alpha},f_{\bar\alpha}]
\\
=&
\frac{1}{L}
\sum_{\substack{k_1^{}k_2^{} \\ k_{1'}k_{2'}\alpha}}
W_{1^{}2^{},{1'}{2'}}
\frac{1}{2}
\big[
E_{k_1\alpha}+
\!\!\!\!\!\!\underbrace{E_{k_1\alpha}}_{\substack{\textrm{Exchange}\ k_1\leftrightarrow k_2,\\ k_{1'}\leftrightarrow k_{2'}\ \textrm{and}\ \alpha\leftrightarrow\bar\alpha}}\!\!\!\!\!\!
\big]
\Big\{
f_{\alpha1'}[1-f_{\alpha1}]f_{\bar{\alpha}2'}[1-f_{\bar{\alpha}2}]
-
f_{\alpha1}[1-f_{\alpha1'}] f_{\bar{\alpha}2}[1-f_{\bar{\alpha}2'}]
\Big\}
\nonumber\\
%%%%%%%%%%
=&
\frac{1}{4L}\!
\sum_{\substack{k_1^{}k_2^{}\\ k_{1'}k_{2'}\alpha}}
W_{1^{}2^{},{1'}{2'}}
\big[
E_{k_1\alpha}+E_{k_2\bar\alpha}+
\underbrace{E_{k_1\alpha}+E_{k_2\bar\alpha}}_{\substack{\textrm{Exchange:}\\ k_1\leftrightarrow k_{1'},\ k_2 \leftrightarrow k_{2'}}}
\big]
\Big\{
f_{\alpha1'}[1-f_{\alpha1}]f_{\bar{\alpha}2'}[1-f_{\bar{\alpha}2}]
-
f_{\alpha1}[1-f_{\alpha1'}] f_{\bar{\alpha}2}[1-f_{\bar{\alpha}2'}]
\Big\}
=0
\nonumber
\end{align}
\end{widetext}
due to energy conservation, $E_{k_1\alpha}+E_{k_2\bar\alpha}-E_{k_{1'}\alpha}-E_{k_{2'}\bar\alpha}=0$, and by using $W_{1^{}2^{},{1'}{2'}}=W_{2^{}1,{2'}{1'}}=W_{{1'}{2'},1^{}2^{}}$ in the summation index exchanges indicated. From this, we obtain the conserved quantity $\sum_{\alpha}Z_{\alpha}$ indicated in Eq.~(\ref{energy-conservation-general}). 

The above considerations can easily be extended to include spin, intra-mode scattering and other types of collision integrals.  Furthermore, \emph{if} we also had momentum conservation, then $\sum_{k\alpha}kv_{k\alpha}f_\alpha(k,x)$ would also be conserved. However, this is not the case here.

\section{Comparison of different initial shapes of the EDIE}\label{app:initial-blobs}

\begin{figure*}
\includegraphics[width=0.8\columnwidth]{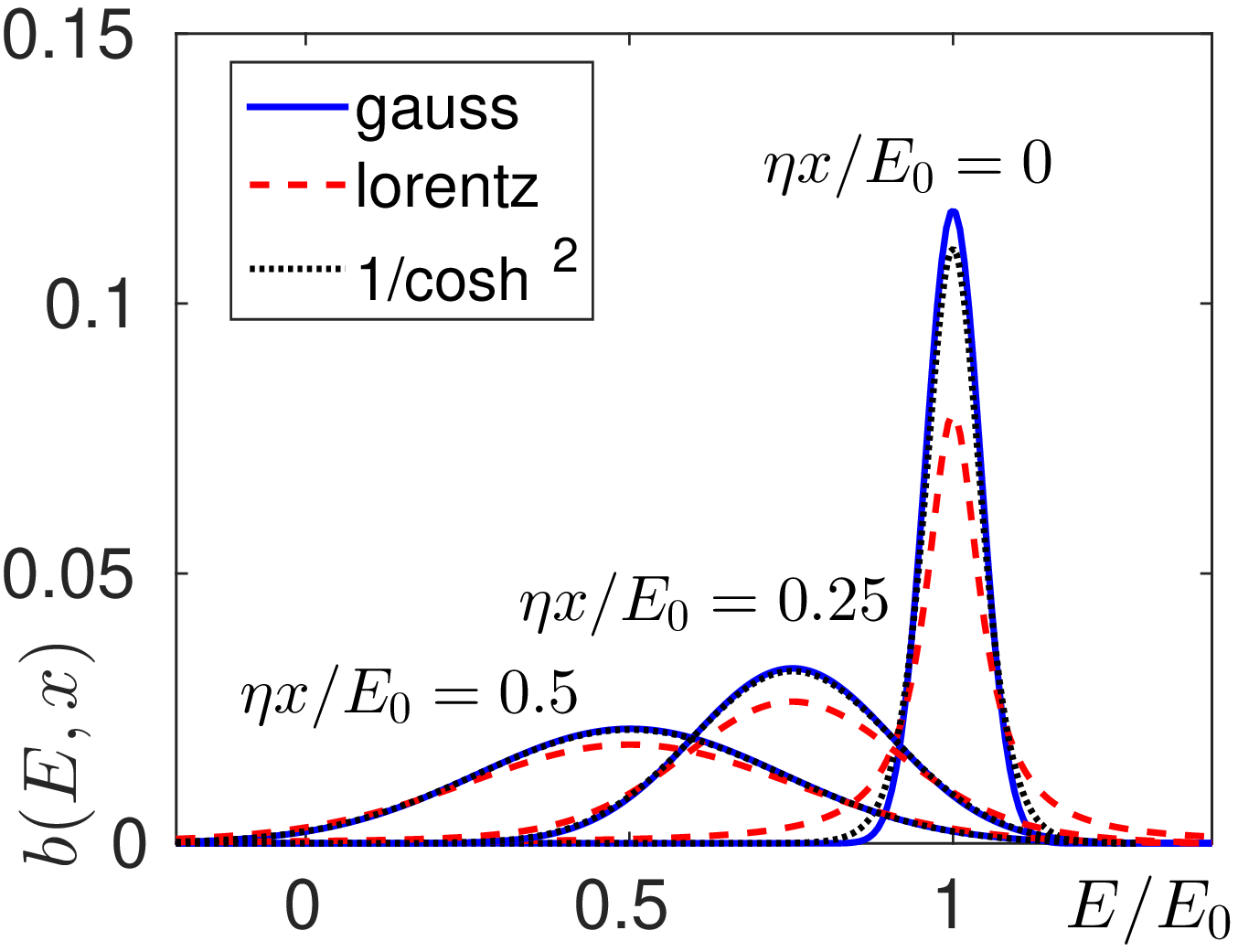}\hspace{1.4cm}
\includegraphics[width=0.8\columnwidth]{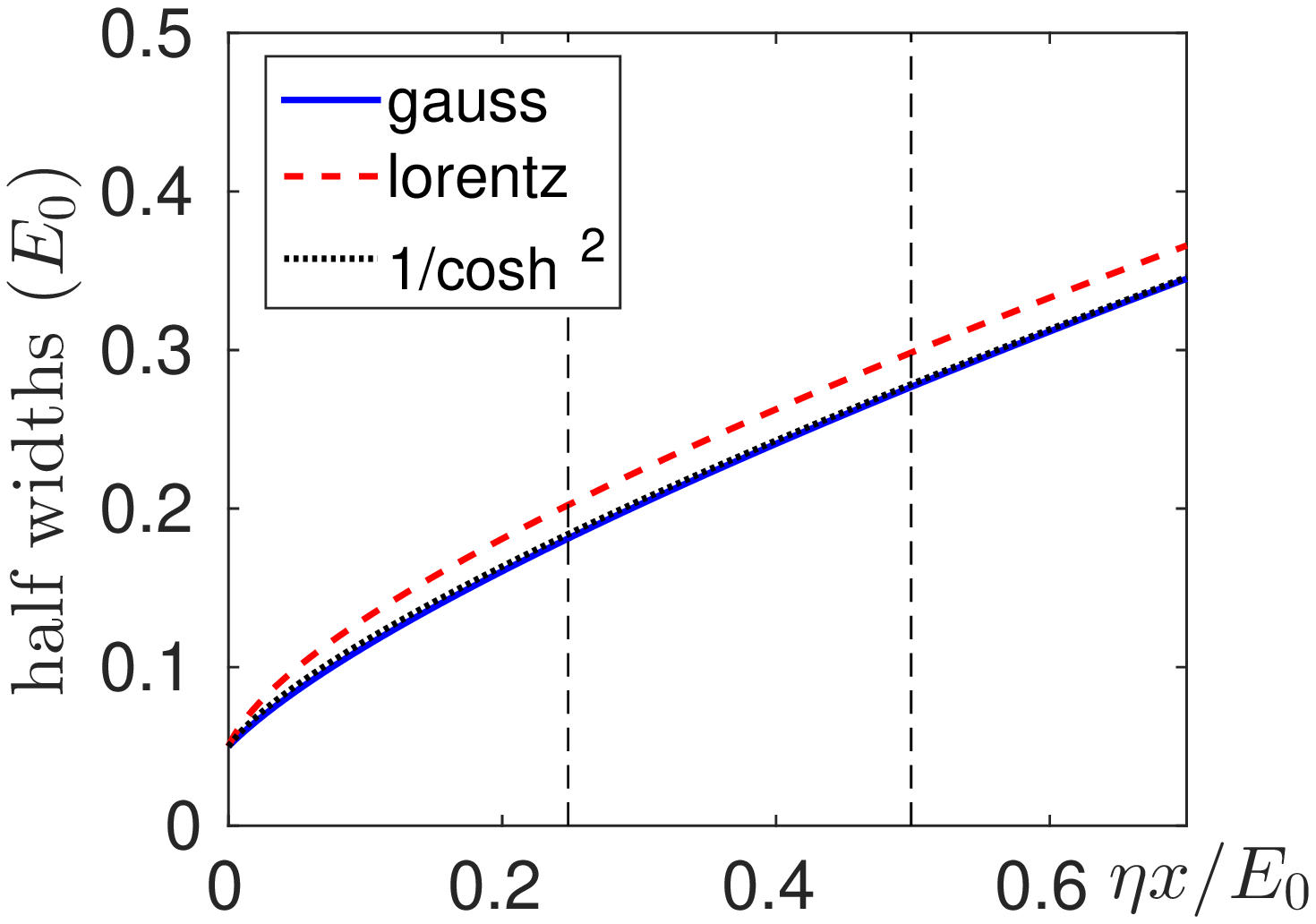}
\caption{(Color online) Comparison of the electron relaxation for different initial shapes of the EDIE. Left panel: The shape of the EDIE at the distances marked by the vertical dashed lines on the right panel. Right panel: The half-widths taken at half of the maximum of $b(E,x)$ versus $x$ in units of $E_0/\eta$. The parameters used are $N_b/\rho_o= 0.5$, $E_0 = 40$, $\Gamma_{\rm l} = 2$ (lorentzian width), $\eta = 0.014$ and $\kb T = 1.25$.}
\label{fig:compare-widths}
\end{figure*}

In Sec.~\ref{sec:coupl-fokk-planck} of the main text, we focused on the relaxation of an initial gaussian EDIE. In this appendix we show that the relaxation is qualitatively similar for different shapes of the initial distribution. Specifically we do this by solving numerically the partial differential equation~(\ref{eq:30}) and computing $D_i(0,x)$ according to Eq.~(\ref{eq:28}) for the following initial distributions
\begin{align}
b_{\rm  l}(E,x=0)&=
\frac{N_b}{\pi\rho_o}\frac{\Gamma_{\rm l}}{\Gamma_{\rm l}^2+(E-E_0)^2},
\\
b_{\rm c}(E,x=0)&=
\frac{N_b}{4\Gamma_{\rm c}\rho_o}\cosh^{-2}\left(\frac{E-E_0}{2\Gamma_{\rm c}}\right),
\\
b_{\rm g}(E,x=0)&=
\frac{N_b}{\Gamma_{\rm g}\sqrt{\pi}\rho_o}\exp\left(-\frac{(E-E_0)^2}{\Gamma_{\rm g}^2}\right).
\end{align}
In order to facilitate the comparison, we choose distributions with equal initial half-widths at half maximum equal to $\Gamma_{\rm l}$, which is achieved by letting $\Gamma_{\rm c}=\Gamma_{\rm l}/(2{\rm arccosh}(\sqrt{2}))$ and $\Gamma_{\rm g}=2\Gamma_{\rm l}/\sqrt{\ln(2)}$. (In the main text, we use the notation $\Gamma_0$ instead of $\Gamma_{\rm g}$ for simplicity.) Only the initial gaussian distribution keeps its functional form for $x>0$.  Fig.~\ref{fig:compare-widths} shows the evolution of the three types of initial distributions as well as their half-widths. The broadening of the distributions is indeed similar. Note however that the initial lorentzian distribution initially broadens somewhat faster than either the gaussian or the $\cosh^{-2}$ distributions. However, the lorentzian distribution also has more weight in its longer power-law tails compared to the exponential tails of the two other initial shapes. Furthermore, since all these initial shapes are symmetric, their maximum evolve identically as shown in Appendix \ref{app:symmetric-packet}.

\section{Evolution of a symmetric EDIE following the Fokker-Planck equation}\label{app:symmetric-packet}

In this Appendix, we discuss the class of EDIE $b(E,x)$ evolving according to the Fokker-Planck equation (\ref{eq:30}) or (\ref{eq:13}), which are initially symmetric (i.e.~even) in energy around their maxima $E_{\textrm{max}}(x)$. We show that (i) the maximum and the average energy are identical for a symmetric EDIE, $\langle E\rangle=E_{\textrm{max}}(x)$, and (ii) an initially symmetric EDIE at $x=0$ remains symmetric for $x>0$. 

We begin by showing $\langle E\rangle=E_{\textrm{max}}(x)$ by using that a symmetric EDIE fulfills $b(E+E_{max}(x),x)=b(-E+E_{max}(x),x)$ for any $x$. A direct calculation yields  
\begin{align} 
\langle E\rangle
\equiv& 
\frac{\rho_o}{N_b}  
\int_{-\infty}^{\infty}d E E b(E,x) 
\nonumber\\
=&
\frac{\rho_o}{N_b}  
\int_{-\infty}^{\infty}d E' [E'+E_{max}(x)] b(E'+E_{max}(x),x) 
\nonumber\\
=&
\frac{\rho_o}{N_b} 
\int_{-\infty}^{\infty}d E' E' \underbrace{b(E'+E_{max}(x),x)}_{\substack{\textrm{even for}\ b(E,x)\ \textrm{symmetric}}} 
\nonumber\\
&+
E_{max}(x)
\frac{\rho_o}{N_b}  \int_{-\infty}^{\infty}d E' b(E'+E_{max}(x),x) 
\nonumber\\
=&
E_{max}(x),
\end{align}
where a new integration variable $E'=E-E_{max}(x)$ was used and we arrived at the desired result.  

However, having this statement for a symmetric EDIE for any $x$, we are faced with a new question: Does an EDIE remain symmetric for $x>0$, if it is initially symmetric at the injection point $x=0$? In other words, do the Fokker-Planck equation preserve the symmetry around the maximum of the EDIE? To answer this, we write the EDIE in a co-moving frame as
\begin{align} 
b(E,x)=\mathfrak{B}(E-E_{max}(x),x),
\label{eq:ansatz-to-get-rid-of-first-term}
\end{align}
where $E_{max}(x)=E_0-\eta x$ and the initial EDIE $b(E,0)$ is assumed to be symmetric around  $E_0$. By inserting this rewriting into the Fokker-Planck equation (\ref{eq:30}), we see its advantage, namely that the drift term is eliminated, i.e. 
\begin{align} 
\mathfrak{B}^{(0,1)}(E-&E_{max}(x),x)=
\eta D_{i}(0,x)\mathfrak{B}^{(2,0)}(E-E_{max}(x),x). 
\label{eq:simplified-FP-for-b}
\end{align}
Here the notation $\mathfrak{B}^{(0,1)}$ means differentiation with respect to the second entry in the function one time and $\mathfrak{B}^{(2,0)}$ means differentiation with respect to the first entry two times etc. Since the differentiation is with respect to the entry in the function (not energy nor space), we can replace $E-E_{max}(x)$ by $\bar{E}$ for convenience, so  
\begin{align} 
\mathfrak{B}^{(0,1)}(\bar{E},x)=\eta D_{i}(0,x)\mathfrak{B}^{(2,0)}(\bar{E},x). 
\end{align}
Now we integrate over space form the injection point $x=0$ to $x$ and use that we know the full functional form of the initial EDIE at $x=0$, i.e.
\begin{align} 
\mathfrak{B}(\bar{E},x)=
\mathfrak{B}(\bar{E},0)
+\eta 
\int_{0}^{x}d x_1 D_{i}(0,x_1)\mathfrak{B}^{(2,0)}(\bar{E},x_1). 
\end{align}
This equation has an iterative structure similar to a Dyson equation, so we can insert it into itself and obtain
\begin{align} 
\mathfrak{B}(\bar{E},x)
=&
\mathfrak{B}(\bar{E},0)
+\eta \left[\int_{0}^{x}d x_1 D_{i}(0,x_1)\right]\mathfrak{B}^{(2,0)}(\bar{E},0) 
\nonumber\\
&+\eta^2 
\int_{0}^{x} \!\!d x_1     D_{i}(0,x_1)
\int_{0}^{x_1}\!\!d x_2 D_{i}(0,x_2)\mathfrak{B}^{(4,0)}(\bar{E},x_2).  
\nonumber
\end{align}
Repeating this procedure, we find a formal solution 
\begin{widetext}
\begin{align} 
\mathfrak{B}(\bar{E},x)=
\mathfrak{B}(\bar{E},0)
+
\sum_{n=1}^{\infty} 
\eta^n
\left[ \int_{0}^{x} d x_1 D_{i}(0,x_1)
\int_{0}^{x_1} d x_2 D_{i}(0,x_2)
\cdots \int_{0}^{x_{n-1}} d x_n D_{i}(0,x_n)\right]
\mathfrak{B}^{(2n,0)}(\bar{E},0),  
\label{eq:formal-solution-for-blob}
\end{align}
\end{widetext}
where $x_0$ should be understood as $x$. From this formal solution, we now argue that $b(E,x)$ remains symmetric for $x>0$, if $b(E,x=0)$ is symmetric. First, we note that if $\mathfrak{B}(\bar{E},x)$ is even in $\bar{E}$, then $b(E,x)$ is symmetric. Therefore, we want to show that $\mathfrak{B}(\bar{E},x)$ is even, if $\mathfrak{B}(\bar{E},x=0)$ is even. If $\mathfrak{B}(\bar{E},x=0)$ is even, then so are all the even derivatives $\mathfrak{B}^{(2n,0)}(\bar{E},0)$ for all $n\in\mathbb{N}$. Thus,   it follows from Eq.~(\ref{eq:formal-solution-for-blob}) that  $\mathfrak{B}(\bar{E},x)$ is indeed even in energy, if $\mathfrak{B}(\bar{E},x=0)$ is even. Hence, we arrive at the desired result, namely, that the EDIE remains symmetric, if it is symmetric initially. 

\section{Various calculational details}

This Appendix serves as a help with various mathematical details in the paper. 

\subsection{Detailed evaluation of $\p_{\omega}D_{i}(0,x)=1/2$}\label{app:dD}

Here we show that $\p_{\omega}D_{i}(0,x)=1/2$ for all $x$. We only use that $f_i(E,x)$ is fully occupied (empty) for very low (high) energy, so knowledge of the entire distribution $f_i$ is not required. From the definition of  $D_{i}(\omega,x)$ in Eq.~(\ref{eq:20}), we obtain
\begin{align}
\p_{\omega}D_{i}(0,x)
&=
\int_{-\infty}^{\infty} d E \big[-\p_Ef_{i}(E,x)\big]\big[1-f_{i}(E,x)\big]\nonumber\\
%%%%
&=-\int_{-\infty}^{\infty}\!\!\! d E \p_Ef_{i}(E,x)
+\int_{-\infty}^{\infty} \!\!\!d E [\p_Ef_{i}(E,x)]f_{i}(E,x)\nonumber\\
%%%%
&=1
+\int_{-\infty}^{\infty} d E [\p_Ef_{i}(E,x)\big]f_{i}(E,x)
%%%%
=\frac{1}{2},
\label{eq:dD-bar-alpha-lig-halv-model1}
\end{align}
since partial integration gives  
\begin{align}
\int_{-\infty}^{\infty} d E [\p_Ef_{i}(E,x)\big]f_{i}(E,x)
=-\frac{1}{2},  
\end{align}
using the high and low energy limits of $f_i(E,x)$ to obtain a  boundary term of $-1$.  The result, $\p_{\omega}D_{i}(0,x)=1/2$,  is a useful and astonishing simplification. The result $\p_\omega\mathfrak{D}_o(0,x)=1/2$ is shown in the same way.

\subsection{Detailed evaluation of $\p_xZ_{b}$, $\p_xZ_{\fo}$ and $\p_xZ_{f_i}$}\label{app:dZ}

For completeness, we show how to evaluate $\p_xZ_{b}$, $\p_xZ_{\fo}$ and $\p_xZ_{f_i}$ in Eq.~(\ref{eq:Zs-model2}) in Sec.~\ref{conserved-quan-model2}. 

First, we consider $\p_xZ_{b}$. We insert the right-hand side of the Fokker-Planck Eq.~(\ref{eq:13}) (or equivalently Eq.~(\ref{eq:30})) into the definition of $Z_{b}(x)$ in Eq.~(\ref{eq:def-Z-g}) and use partial integration, i.e.
\begin{align}
\p_xZ_{b}(x)
&= \int^{\infty}_{-\infty}d E (E-\bar{\mu}_{b}) \p_x b(E,x)
\nonumber\\ 
%%%%
&=\eta
\int^{\infty}_{-\infty}d E (E-\bar{\mu}_{b})
\big[\p_Eb(E,x)+
D_{i}(0,x)
\p^2_Eb(E,x)\big] 
\nonumber
\\
&=\eta 
\Bigg\{
\bigg[(E-\bar{\mu}_{b})
\big[b(E,x)+D_{i}(0,x)\p_Eb(E,x)\big]
\bigg]^{E\rightarrow\infty}_{E\rightarrow-\infty}
\nonumber\\
&
\hspace{6mm}
-\int^{\infty}_{-\infty}d E \big[b(E,x)
+D_{i}(0,x)\p_Eb(E,x)\big]
\Bigg\} 
%%%%%
\nonumber\\
&=-\eta \frac{N_b}{\rho_o}. 
\end{align} 
Here we use that $b(E,x)$ and $\p_Eb(E,x)$ goes to zero faster than $\vert E-\bar{\mu}_{b}\vert$ increases for $E\rightarrow \pm \infty$ as well as $\int d Eb(E,x)=N_b/\rho_o$. Interestingly, this calculation shows explicitly that the diffusion term $D_{i}(0,x)\p^2_Eb$ of the Fokker-Planck equation for the EDIE does not provide a source for energy redistribution between the ESs.

Secondly, we find $\p_xZ_{\fo}$ by inserting the right-hand side of Eq.~(\ref{eq:14}) into the definition of $Z_{\fo}$ and use partial integration such that
\begin{align}
\p_xZ_{\fo}(x)
&=\eta
\int^{\infty}_{\epsilon_c}d E
(E-\bar\mu_{\fo})
\nonumber\\ 
&\hspace{3mm}
\times
\Big\{
\p_E\big\{[1-\fo(E,x)]\fo(E,x)\big\}
+D_{i}(0,x)\p_E^2\fo(E,x)\Big\}
\nonumber\\ 
&=\eta 
\Bigg\{
\Big[
(E-\bar\mu_{\fo})[1-\fo(E,x)]\fo(E,x)
\Big]^{E\rightarrow\infty}_{E\rightarrow\epsilon_c}
\nonumber\\ 
&\hspace{5mm}
-\int^{\infty}_{\epsilon_c}d E[1-\fo(E,x)]\fo(E,x)
\nonumber\\
&\hspace{5mm}
+
\Big[(E-\bar\mu_{\fo})D_{i}(0,x)\p_E\fo(E,x)
\Big]^{E\rightarrow\infty}_{E\rightarrow\epsilon_c}
\nonumber\\ 
&\hspace{5mm}
- 
D_{i}(0,x)
\int^{\infty}_{\epsilon_c}d E\p_E\fo(E,x)
\Bigg\} 
\nonumber\\
&=-\eta \mathfrak{D}_{o}(0,x)
+\eta D_{i}(0,x).
\end{align} 
Here we used the rewriting $[1-2\fo(E,x)]\p_E\fo(E,x)=\p_E\big\{[1-\fo(E,x)]\fo(E,x)\big\}$ in the drift-term of the Fokker-Planck equation, the fact that both $[1-\fo]\fo$ and $\p_E\fo$ goes to zero faster than $\vert E-\bar{\mu}_{\fo}\vert$ for high and low energies, the definition of $\mathfrak{D}_{o}(0,x)$ in Eq.~(\ref{eq:def-Dfrak}) and that $\int d E\p_E\fo(E,x)=-1$. Finally, $\p_xZ_{f_i}$ is found essentially by combining the arguments leading to $\p_xZ_{b}$ and $\p_xZ_{\fo}$  above. 

\section{Approximate solutions of the effective temperatures}\label{app:asym-eff-temp}

\begin{figure}
\includegraphics[width=1\columnwidth]{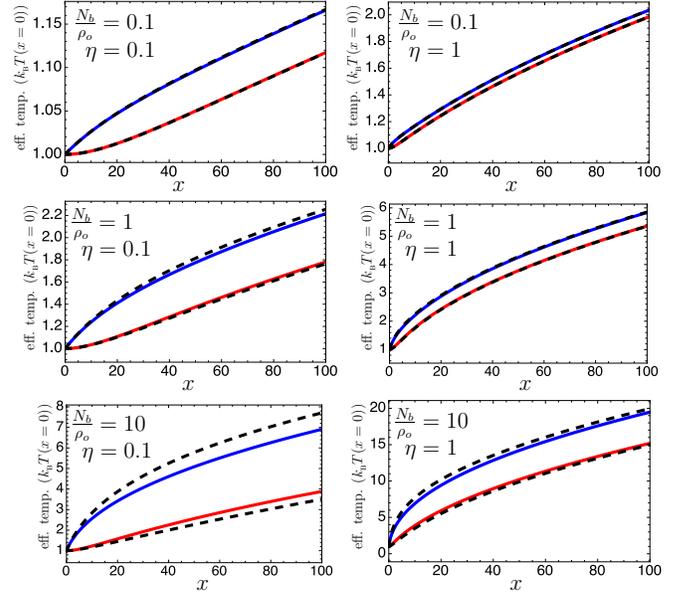}
\caption{(Color online) Comparison of the effective temperatures found numerically from Eqs.~(\ref{eq:eff-temp-diff-eq}) ($\kb T_i(x)$ blue and $\kb T_o(x)$ red) and the approximations in Eqs.~(\ref{eq:Ts-approx}) and (\ref{eq:Td-approx}) (dashed black curves), all in units of $\kb T$. We vary $N_b/\rho_o$ and $\eta$ and keep $\kb T=1$ (in arbitrary units) fixed. The size of the effective interaction $\eta$ decides how rapidly the intermediate regime with $\kb T_i-\kb T_o\simeq N_b/(2\rho_o)$ is reached. The approximate solutions are seen to work rather well, especially for larger $x$ where $T_d/T_s$ decrease. For instance, if the bottom figures were shown in a larger range of $x$, then the comparison would appear better.  }
\label{fig:approx-eff-temp}
\end{figure}

In this Appendix, we find an approximate solution to the differential equations (\ref{eq:eff-temp-diff-eq}) for the effective temperatures in Sec.~\ref{sec:effect-temp-appr}. The approximation builds on the observation that the sum of the effective temperatures is much larger than their difference. The sum and difference of the effective temperatures are
\begin{align} 
T_s(x)&=\frac{1}{2}[T_i(x)+T_o(x)],
\quad
T_d(x)=\frac{1}{2}[T_i(x)-T_o(x)].
\end{align} 
The evolution of $T_{s}$ and $T_{d}$ are found from Eqs.~(\ref{eq:eff-temp-diff-eq}) to be
\begin{subequations} 
\label{eq:diff-eq-for-Ts-and-Td}
\begin{align} 
\kb \p_x T_s(x)
&=
\frac{3\eta}{2\pi^2}
\Bigg[
\frac{N_b/(\rho_o\kb)-2T_d}{T_d+T_s}
+\frac{2 T_d}{T_s-T_d}
\Bigg],
\label{eq:diff-eq-for-Ts}
\\
\kb \p_x T_d(x)
&=
\frac{3\eta}{2\pi^2}
\Bigg[
\frac{N_b/(\rho_o\kb) (T_d -  T_s) + 4  T_d T_s}{ T_d^2 -  T_s^2}
\Bigg]. 
\label{eq:diff-eq-for-Td}
\end{align} 
\end{subequations} 
The effective temperatures of the two ESs, $T_o(x)$ and $T_i(x)$, will grow for increasing $x$ as discussed below Eqs.~(\ref{eq:eff-temp-diff-eq}). Moreover, the exact relation (\ref{eq:cons-law-eff-temp}) shows that the sum of the squared effective temperatures also grows. Therefore, we construct approximate solutions relaying on $T_s(x)$ being much larger than $T_d(x)$ as $x$ increases.  For $T_s(x)\gg T_d(x)$, Eq.~(\ref{eq:diff-eq-for-Ts}) for $T_s$ simplifies to 
\begin{align} 
\kb \p_x T_s(x)
&\simeq
\frac{3\eta}{2\pi^2}
\frac{N_b/\rho_o}{\kb T_s},
\end{align} 
which has the solution 
\begin{align}  
\kb T_s(x)\simeq\sqrt{3 \eta  xN_b/(\rho_o\pi^2)+(\kb T)^2}, 
\label{eq:Ts-approx}
\end{align} 
using that $T_i(0)=T_o(0)=T$. Similarly, for $T_s(x)\gg T_d(x)$  Eq.~(\ref{eq:diff-eq-for-Td}) for $T_d$ becomes
\begin{align} 
\kb \p_x T_d(x)
&\simeq
\frac{3\eta}{2\pi^2}
\Bigg[
\frac{N_b/(\rho_o\kb) - 4  T_d}{T_s}
\Bigg],
\end{align} 
Inserting the solution of $T_s$ and using $T_d(0)=0$, we obtain 
\begin{align} 
\label{eq:Td-approx}
\kb T_d(x)
&\simeq
\frac{N_b}{4\rho_o}
\Bigg[
1-
e^{\frac{4\kb T}{N_b/\rho_o}-4 \sqrt{\frac{3 \eta  x}{\pi^2N_b/\rho_o}+\Big[\frac{\kb T}{N_b/\rho_o}\Big]^2}}
\Bigg].
\end{align} 
The approximate effective temperatures are now easily found from $T_i=T_s+T_d$ and $T_o=T_s-T_d$. Since the exponential term vanish for $x\gg N_b/(\rho_o\eta)$, we find the difference of the effective temperatures to approach a simple constant, 
\begin{align} 
\kb T_d(x)\simeq \frac{N_b}{4\rho_o}
\end{align} 
in the regime of intermediate distances, $N_b/(\rho_o\eta)\ll x\ll\ell_{\textsc{fp}}$, as given in Eq.~(\ref{eq:approx-eff-temp-text}) of the main text. The upper bound $\ell_{\textsc{fp}}$ is the limit of the Fokker-Planck approach and the intermediate regime is only reached for a sufficiently large injection energy $E_0$. Our approximations fit well with the numerical examples shown in Figs.~\ref{fig:iter_eff_temp2} and \ref{fig:compare-eff-temp-model2} and a direct comparison is found in Fig.~\ref{fig:approx-eff-temp} and its caption. Both the numerical and approximate solutions agree with a decreasing $T_d(x)/T_s(x)$ for large $x$. 

\section{On the comparison of the effective temperature approach, the coupled Fokker-Planck equations (\ref{eq:Full-FP-model2}) and the full kinetic equations~(\ref{full-KE})}\label{app:compare-eff-temp-and-model2}

\begin{figure}
\includegraphics[width=0.85\columnwidth]{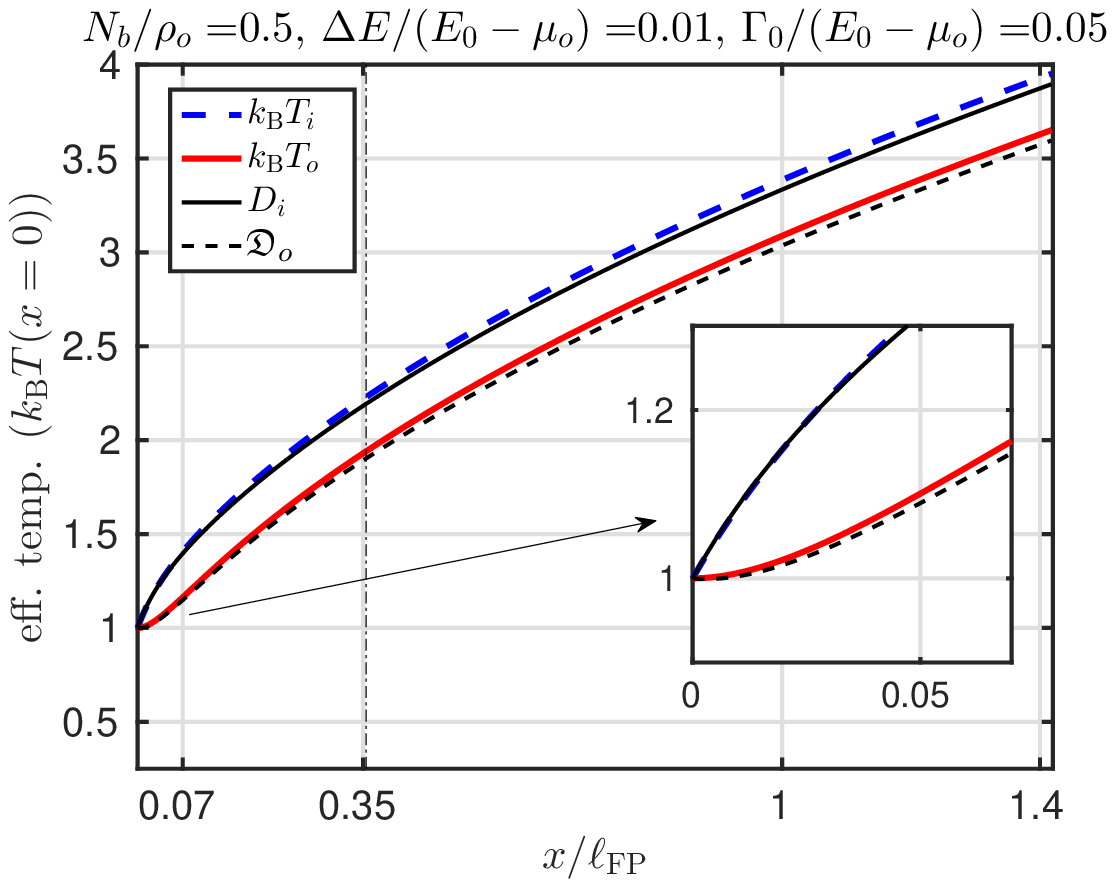}\vspace{2mm}
\includegraphics[width=0.83\columnwidth]{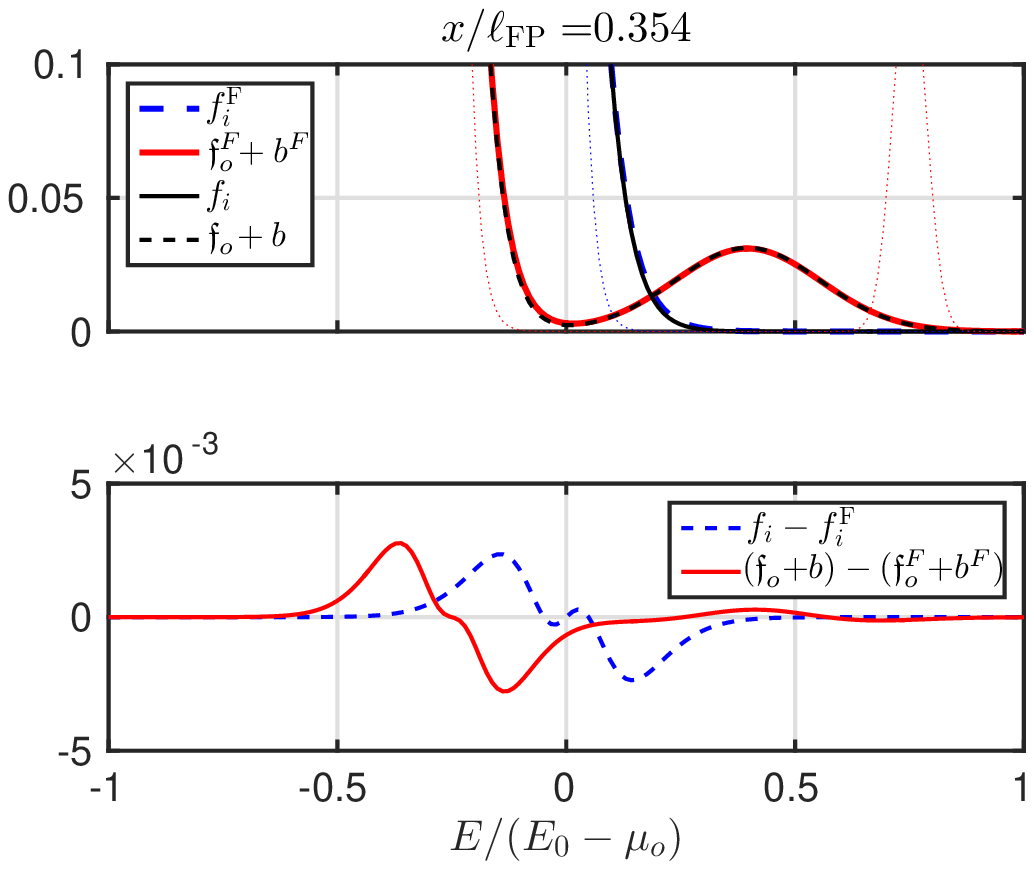}
\caption{(Color online) Comparison of the effective temperature approach in Sec.~\ref{sec:effect-temp-appr} and a numerical solution of the coupled Fokker-Planck Eqs.~(\ref{eq:Full-FP-model2}). The effective temperatures $\kb T_i(x)$ and $\kb T_o(x)$ are compared to the energy smearing of the Fermi levels, $D_i(0,x)$ and $\mathfrak{D}_o(0,x)$, from the Fokker-Planck Eqs.~(\ref{eq:Full-FP-model2}) in the top panel, while a snapshot of the distributions are compared in the bottom panel. We use the same nomenclature and parameters as in Figs.~\ref{fig:iter_eff_temp2} and \ref{fig:distr_iter} in the main text, except for a five times larger EDIE, i.e. $N_b/\rho_o=0.5$. Note that neither the effective temperatures nor the energy smearing of the Fermi levels saturate for $x>\ell_{\textsc{fp}}$, since these were only created to capture the physics of a EDIE separated from the Fermi sea, i.e.~$x<\ell_{\textsc{fp}}$.}
\label{fig:compare-eff-temp-model2}
\end{figure}

The effective temperature approach in Sec.~\ref{sec:effect-temp-appr} is based directly on the coupled Fokker-Planck Eqs.~(\ref{eq:Full-FP-model2}) including both Fermi seas. Therefore, we compare these two approaches here for completeness in Fig.~\ref{fig:compare-eff-temp-model2}. In the main text, the effective temperature approach is compared to the full kinetic Eqs.~(\ref{full-KE}) instead, see Figs.~\ref{fig:iter_eff_temp2} and \ref{fig:distr_iter}. 

We find that the effective temperature approach agrees almost perfectly with a numerical solution of the coupled Fokker-Planck Eqs.~(\ref{eq:Full-FP-model2}) for the parameters of Figs.~\ref{fig:iter_eff_temp2} and \ref{fig:distr_iter} in the main text, i.e. $N_b/\rho_o=0.1$ relevant for a small EDIE. However, if the EDIE is larger, e.g. $N_b/\rho_o=0.5$ as in Fig.~\ref{fig:compare-eff-temp-model2}, then the comparison to the Fokker-Planck Eqs.~(\ref{eq:Full-FP-model2}) is still rather good whereas it appears to compare less well to the full kinetic Eqs.~(\ref{full-KE}). The difference between these two comparisons is \emph{not} due to an actual difference between the Fokker-Planck Eqs.~(\ref{eq:Full-FP-model2}) and the full kinetic Eqs.~(\ref{full-KE}). Instead it is an artefact of the simple way we extract the energy smearing of the outer ES's Fermi level from the full kinetic Eqs.~(\ref{full-KE}) in Figs.~\ref{fig:iter_eff_temp2} and \ref{fig:distr_iter}. In these figures, we use $D_o(0,x)-N_b/\rho_o$ as the smearing of the outer ES's Fermi level, since the full kinetic Eqs.~(\ref{full-KE}) do not give direct access to the Fermi sea part of the distribution separated from the EDIE. However, if we wanted to be more precise, then we could compare $\kb T_o(x)$ to $D_o(0,x)-N_b/\rho_o+N_b^2/(\rho_o^2\Gamma(x)\sqrt{2\pi})$, since 
\begin{align} 
D_o(0,x)&=
\int^\infty_{-\infty} dE \; [\fo+b][1-\fo-b]
\nonumber\\
&=\int^\infty_{-\infty} dE \; \Big( \fo[1-\fo] + b- b^2-2b\fo \Big)
\nonumber\\
&=\frak{D}_o(0,x) + 
\frac{N_b}{\rho_o}- \frac{N_b^2}{\rho_o^2\Gamma(x)\sqrt{2\pi}}
-2\int^\infty_{-\infty} \!dEb\fo
\end{align} 
using a Gaussian EDIE $b$. For small $x$, the cross term, $b\fo$, vanish whereas the $b^2$ term remains finite, yet suppressed as $N_b^2/(\rho_o^2\Gamma(x))$. In fact, taking the $b^2$ term into account, the small difference between $\kb T_o(x)$  and  $D_o(0,x)-N_b/\rho_o$ for small $x$ in the inset of Fig.~\ref{fig:iter_eff_temp2} disappears.  

In contrast to the artefact of the comparison discussed above, the small difference between the effective temperature approach and the coupled Fokker-Planck Eqs.~(\ref{eq:Full-FP-model2}) seen on Fig.~\ref{fig:compare-eff-temp-model2} is indeed an actual difference between the two approaches.

%\bibliography{Ref-blob-paper-2013}

\begin{thebibliography}{70}
\bibitem[{foo({\natexlab{d}})}]{footnote-index-specification}
\bibinfo{note}{The ES subscript $\alpha$ on
  $I_{k_1^{}x\alpha}[f_\alpha,f_{\bar\alpha}]$ is in some sense redundant,
  since the first distribution function in the square bracket \emph{is forced
  to have the same} ES index as the subscript. We keep the subscript here,
  however, because $\alpha$ does not only enter
  $I_{k_1^{}x\alpha}[f_\alpha,f_{\bar\alpha}]$ through the distribution
  function and since later we will use distribution functions without an ES
  index.}
\bibitem[{foo({\natexlab{e}})}]{footnote-trans-inv-int}
\bibinfo{note}{In contrast, any translational invariant interaction,
  $V(x,x')=V(x-x')$, automatically leads to momentum conserving scattering
  only, since $\langle k_{1'}\alpha,k_{2'}\bar\alpha|V(x-x')|k_{1}^{}\alpha,
  k_{2}^{}\bar\alpha\rangle\propto \delta_{k_{1'}+k_{2'},k_{1}^{}+k_{2}^{}}$.}
\bibitem[{foo({\natexlab{h}})}]{footnote-T_inf-lower-than-T}
\bibinfo{note}{Here, we argue why the Fermi sea and the EDIE have a large
  overlap in energy for $E_0-\mu_o <N_b/(2\rho_o)$ using a Gaussian EDIE
  Eq.~(\ref{eq:gauss-EDIE-model2}). The entire initial distribution function of
  the outer ES has to fulfil $0\leq\fo(E,0)+b(E,0)\leq1$ for all $E$. In
  particular for $E=E_0$, we have
  $\fo(E_0,0)+N_b/(\Gamma_0\rho_o\sqrt{\pi})\leq1$ such that
  $N_b/(\Gamma_0\rho_o\sqrt{\pi})<1$ at least should be fulfilled (although it
  is not a sufficient condition). Using these inequalities, we can compare the
  full-width at half maximum of the EDIE, $2\Gamma_0\ln (2)$, to the energy
  difference between the maximum of the EDIE and the Fermi level, i.e.
  $(E_0-\mu_o)/(2\Gamma_0\ln (2)) <N_b/(4\ln
  (2)\rho_o\Gamma_0)<\sqrt{\pi}/(4\ln (2))\simeq 0.64$. Hence, the EDIE has a
  large overlap with the Fermi sea.}
\end{thebibliography}
%\bibliographystyle{apsrev}

\end{document}